\documentclass[aps,prb,reprint,superscriptaddress]{revtex4-2}
\usepackage{graphicx}

\begin{document}

\title[Atomic relaxation and flat bands in strain engineered TMD moir\'{e} systems]{Atomic relaxation and flat bands in strain engineered transition metal dichalcogenide bilayer moir\'{e} systems}
\author{Sudipta Kundu $^*$}
\affiliation{Centre for Condensed Matter Theory, Department of Physics, Indian Institute of Science, Bangalore, India}

\author{Indrajit Maity $^*$}
\affiliation{Centre for Condensed Matter Theory, Department of Physics, Indian Institute of Science, Bangalore, India}
\def\thefootnote{*}\footnotetext{These authors contributed equally}\def\thefootnote{\arabic{footnote}}
\author{Robin Bajaj}
\affiliation{Centre for Condensed Matter Theory, Department of Physics, Indian Institute of Science, Bangalore, India}

\author{H. R. Krishnamurthy}
\affiliation{Centre for Condensed Matter Theory, Department of Physics, Indian Institute of Science, Bangalore, India}
\affiliation{International Centre for Theoretical Sciences, Tata Institute of Fundamental Research, Bengaluru, India}
\author{Manish Jain}
\affiliation{Centre for Condensed Matter Theory, Department of Physics, Indian Institute of Science, Bangalore, India}

\begin{abstract}
Strain-induced lattice mismatch leads to moir\'{e} patterns in homobilayer
transition metal dichalcogenides (TMDs).  We investigate the structural and
electronic properties of such strain-induced moir\'{e} patterns in TMD homobilayers. These moir\'{e} patterns
 consist of several stacking domains which are separated by 
{\em tensile} solitons. Relaxation of these systems distributes the strain unevenly
in the moir\'{e} superlattice, with the maximum strain energy concentrating
at the highest energy stackings. The order parameter distribution shows the
formation of {\em aster} topological defects at the same sites.
In contrast, {\em twisted} TMDs host {\em shear} solitons at the domain walls, and the
order parameter distribution in these systems shows the formation of {\em vortex} defects.
The strain-induced moir\'{e} systems also show the emergence of several well-separated flat
bands at both the valence and conduction band edges, and we observe a
significant reduction in the band gap.  These flat bands in these strain-induced
moir\'{e} superlattices provide platforms for studying the Hubbard model on a
{\em triangular} lattice as well as the {\em ionic} Hubbard model on a {\em honeycomb} lattice.
Furthermore, we study the localization of the wave functions corresponding to
these flat bands.  The wave functions localize at different stackings compared 
to twisted TMDs, and our results on the localization of flat bands at the conduction band edge are in excellent agreement with spectroscopic experiments \cite{strainmose2}.

\end{abstract}

\maketitle

Two perfectly aligned layers of identical two-dimensional (2D) materials have
the same structural periodicity as the individual layers.  The structural
periodicity of the bilayer can be seamlessly manipulated by twisting or
straining one layer relative to another other, or by replacing one of the layers with a different 2D material. All of
these approaches give rise to large-scale moir\'{e} patterns. The periodicities
of the resulting commensurate moir\'e patterns can range from a few nanometers to
hundreds of nanometers. Such moir\'{e} patterns have been shown to significantly alter the
electronic and optical properties of the separate layers - leading, for example, to the emergence of the
correlated electronic states~\cite{graphene2,expt2,wse2_expt2}, superconductivity~\cite{graphene1}, intriguing topological phases~\cite{rubio3,macdonald3,zeng2023thermodynamic}, magnetism~\cite{fu4}, and new types of excitons~\cite{expt1,expt3,expt4,expt5,expt6,expt7,expt8,leroy2,naik2022intralayer,xiong2023correlated,brotons2024interlayer}.

The formation of flat-electronic bands due to the moir\'{e} pattern is responsible for these exotic
properties.  Moir\'{e} patterns formed by twisting transition metal
dichalcogenides (TMDs) layers provide a unique material platform where the
electronic flat bands appear for a broad range of twist angles 
\cite{wse2_expt2,naik1,naik2,continm,sds1,fu1,maity1,maity2,rubio,gammavalley,kundu1,wse2_expt1,lado1,expt9,expt13,rubio2,macdonald1,vitale}.
Most of the studies so far focus on the emergence of flat-electronic bands due to the
moir\'{e} pattern formation from twisting and stacking two dissimilar
materials. Previous experiments, however, suggest that electrons can be efficiently trapped in
strain-engineered moir\'{e} pattern in TMDs \cite{strainmose2}. This hints at
the formation of flat-electronic bands in such systems. Therefore, it is important to explore the possibility of flat electronic bands across a broad range of strains, and the qualitative resemblance of the flat bands and their associated wave functions to those found in the twisted bilayer of TMDs.

Few theoretical studies based on continuum models explored strain-induced moire patterns in bilayer graphene \cite{sinner2023strain,engelke2023topological,escudero2024designing}. A general study on the impact of hetero-strain in 2D materials was reported \cite{kogl2023moire}
. In the strained bilayer of $\mathrm{WSe_{2}}$ , the formation of flat bands has been found from a continuum model-based study \cite{strain1}. 
However, the impact of
structural relaxations on the flat-band formation was not taken into account in this study.
The impact of structural relaxations on twisted bilayer $\mathrm{WSe_{2}}$
including strain, was reported in another study \cite{strain2}. Nonetheless, a
detailed {\em ab initio} study on the formation of flat bands due to {\em only strain engineering} is
missing. Similar to twisted bilayers of TMDs, strain-engineered moir\'{e} patterns may provide a rich platform for manipulating structural and electronic properties, well deserving of further exploration.

In this paper, we investigate the structural and electronic properties of strain-engineered moir\'{e} patterns constructed with bilayer TMDs.  In particular, we consider moir\'{e} patterns formed by applying isotropic biaxial strains ranging between 4\% and 2\%  on the bottom layer of the bilayer TMDs. We find that large atomic relaxations lead to the formation of domain walls connected by topological point defects. Specifically, the domain walls are tensile solitons and the defects are aster topological point defects in the strain-engineered moir\'{e} patterns. The electronic structure shows flat bands near the valence band and the conduction band edges. These flat valence bands can arise from both the $\Gamma$ valley and the $K$ valley depending on the strength of the spin-orbit coupling of the constituent TMDs. The isolated flat bands derived from the $\Gamma$ valley near the valence band edge are ideal for studying the {\em ionic} Hubbard \cite{ionichubbard0,ionichubbard1,ionichubbard2,ionichubbard3,ionichubbard4,ionichubbard5,ionichubbard6} and Hubbard models \cite{hubbard1,hubbard2} on a honeycomb lattice and triangular lattice, respectively. Our findings on the localization of the flat band wavefunctions at the conduction band edge agree with scanning tunneling spectroscopic measurements \cite{strainmose2}. We also extensively compare our results to those of twisted bilayer TMDs, studied previously\cite{naik1}, and find important qualitative differences both at the level of structural and electronic properties. Furthermore, we fit the bands of the moir\'e superlattice (MSL) induced by biaxial strain to tight-binding models and a $k.p$ continuum model (see supplemental material (SM)), wherein we find the parameters to be strain dependent. 
The obtained parameters provide some idea about the probable electronic phases and 
can be used further
to study strongly correlated physics in strain-induced MSLs.

\textit{Methodology}: The moir\'{e} patterns we study are generated using a strained bottom layer and an unstrained top layer. To generate isotropicaly strained moir\'{e} patterns, we apply an equal tensile strain along both the primitive lattice vectors. The moir\'{e} length depends on the
applied strain. The smaller the external strain, the larger is the moir\'{e}
length. The strain-induced MSLs considered here contain thousands of atoms per supercell. We
use an efficient multiscale computational approach for structural optimization of these systems. We
use a classical forcefield to relax the system employing the LAMMPS package 
\cite{lammps}. The intralayer and interlayer interactions are described by a
Stillinger-Weber potential \cite{sw} and a Kolmogorov-Crespi potential 
\cite{kc_naik} respectively.
The electronic structures of the relaxed systems are computed using density functional theory 
as implemented in the SIESTA package 
\cite{siesta}. For details on the calculation parameters, please see SM. 

\begin{figure}[!ht]
\centering
\includegraphics[scale=0.3]{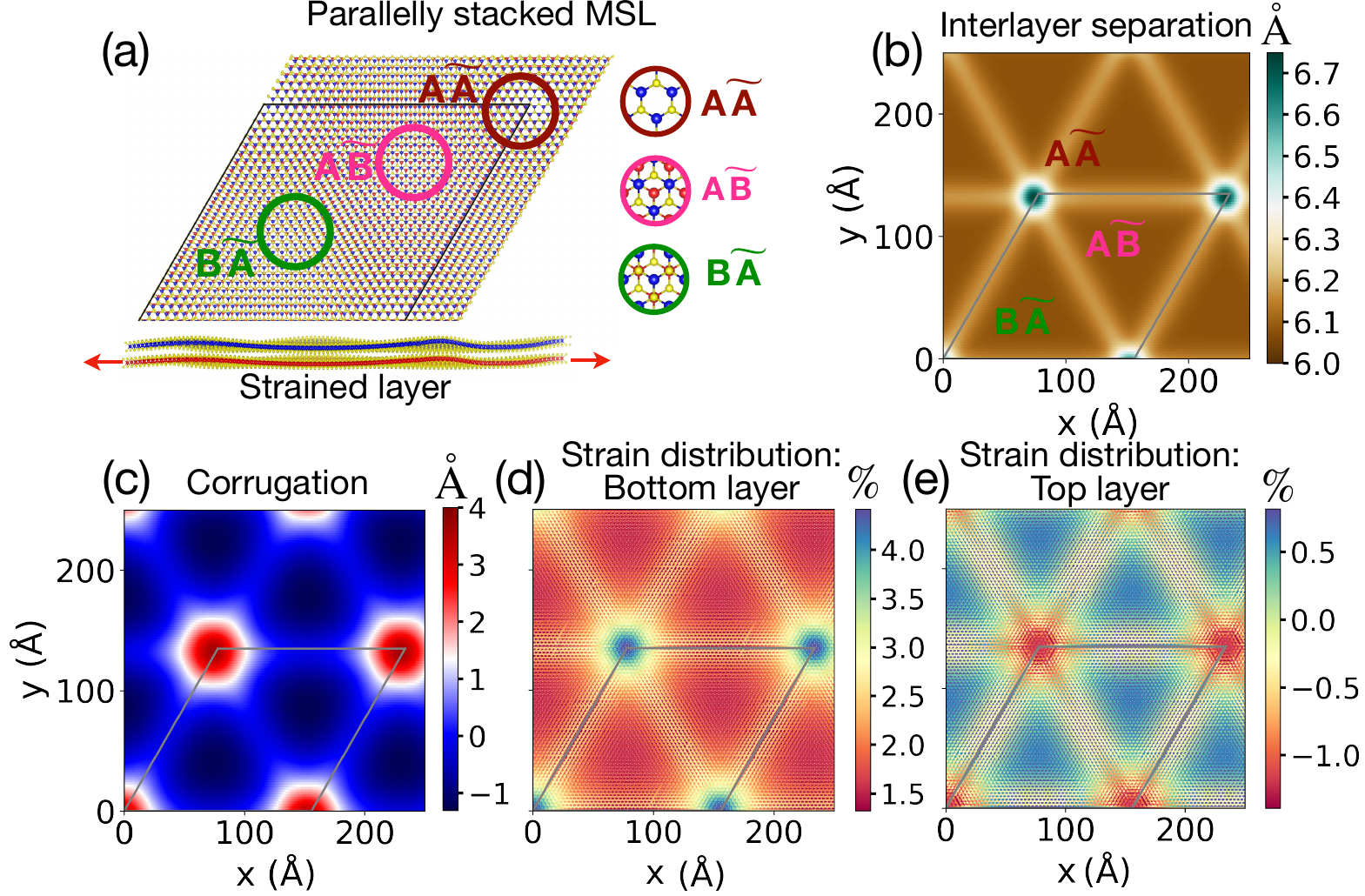}
\caption{\label{fig1}
(a): Top and side views of the relaxed atomic structure of the moir\'{e} pattern formed by applying isotropic biaxial strain to the bottom layer of AA stacked bilayer MoS$_{2}$. The Mo atoms of the bottom and top layers are represented by red and blue colors, respectively. We use yellow for the S atoms of both layers. The circles denote the high-symmetry stackings. The arrows in the side view indicates the applied tensile strain to the bottom layer. (b) and (c): Interlayer separation and corrugation distribution of a 2\% - strain-induced moir\'{e} pattern of MoS$_{2}$. (d) and (e): Distribution of strains in the bottom and top layers of the same moir\'{e} pattern.}
\end{figure}

\textit{Structural relaxation in strain-induced MoS$_2$ MSL with parallel stacking:}
We first discuss the structural aspects of the moir\'{e} pattern obtained by isotropically stretching the bottom layer of a parallelly stacked MoS$_2$ bilayer.  The resulting moir\'{e} pattern consists of $A\widetilde{A}$ (metal (M) on top of $\widetilde{\textrm{M}}$ and chalcogen (X) on top of $\widetilde{\textrm{X}}$),
$A\widetilde{B}$ (Bernal stacking with M on top of $\widetilde{\textrm{X}}$) and $B\widetilde{A}$ (Bernal stacking with X on top of $\widetilde{\textrm{M}}$) high symmetry stackings. We distinguish between the two inequivalent layers by labeling the stackings in the bottom strained layer as $\widetilde{A}$ or
$\widetilde{B}$. Similarly, the atoms of bottom layer are represented by $\widetilde{\textrm{M}}$ or $\widetilde{\textrm{X}}$. Note that the layers are equivalent for moir\'{e} patterns constructed with a twist between the layers.  Conventionally, the high-symmetry-stackings of the twist-induced bilayer are referred to as $AA$, $AB$ and $BA$, respectively~\cite{naik1}.

Owing to the variation of local stacking energy, strain-induced MSLs  show large atomic relaxation similar to those observed in twisted bilayer TMDs \cite{naik2}. The magnitude of the atomic relaxation is determined by a competition between the stacking energy gain and the elastic energy cost. Fig.~\ref{fig1}(a) shows a relaxed biaxially strained bilayer MoS$_2$ for which the bottom layer has been stretched equally along its two primitive lattice vector directions by 2\%, resulting in a MSL containing 14703 atoms. Figs.~\ref{fig1}(b) and (c) show the interlayer spacing and the corrugation to demonstrate the impact of structural relaxation along the out-of-plane direction. 
The interlayer spacing is defined as ($z_{1}-z_{2}$), where $z_1$ and $z_2$ denote the local displacement of each layer along the $z$-direction. Fig.~\ref{fig1}(b) clearly shows the expansion of the favorable $A\widetilde{B}$ and $B\widetilde{A}$ stacking regions and the compression of the unfavorable $A\widetilde{A}$ stacking regions. Note that both $A\widetilde{B}$ and $B\widetilde{A}$ stacking regions form equilateral triangles with same areas. Although the external tensile strain breaks the layer symmetry, the energies of the $A\widetilde{B}$ and $B\widetilde{A}$ stackings are the same, and hence they occupy equal areas in the relaxed moir\'{e} pattern. However, the breaking of layer symmetry leads to a difference in onsite energy, whose impact on electronic structure is discussed later. On the other hand, the in-phase movements of both the layers gives rise to corrugation, 
given by the average movement of both layers,$(\frac{z_{1}+z_{2}}{2})$. The $A\widetilde{A}$ region buckles in an opposite direction to the buckling of $A\widetilde{B}$ and $B\widetilde{A}$. The large corrugation at $A\widetilde{A}$ is connected to the strain redistribution in the MSL. We find qualitatively similar features by varying the biaxial strain for a range of 2\%-3.3\% and comparing the interlayer spacing and corrugations of the relaxed structures. 

Fig.~\ref{fig1}(d) (bottom layer) and (e) (top layer) show the in-plane strain distribution for the same system. It is evident that the 2\% tensile strain applied to the bottom layer is redistributed as a result of atomic relaxation. In particular, the strain transfer leads to a reductions in the local strain at the favorable $A\widetilde{B}$ and $B\widetilde{A}$ stackings and an enhancement at the unfavorable $A\widetilde{A}$ stacking. The domain walls accommodate an intermediate strain. In the top layer, tensile strain appears at the low energy stacking at the expense of compressive strain at the $A\widetilde{A}$ stacking. However, the average strain of the top layer remains zero. Such strain redistribution in both layers lead to larger lattice mismatch at $A\widetilde{A}$ stackings and the highest strain, resulting in larger out-of-plane rearrangement too.

\begin{figure}
\centering
\includegraphics[scale=0.4]{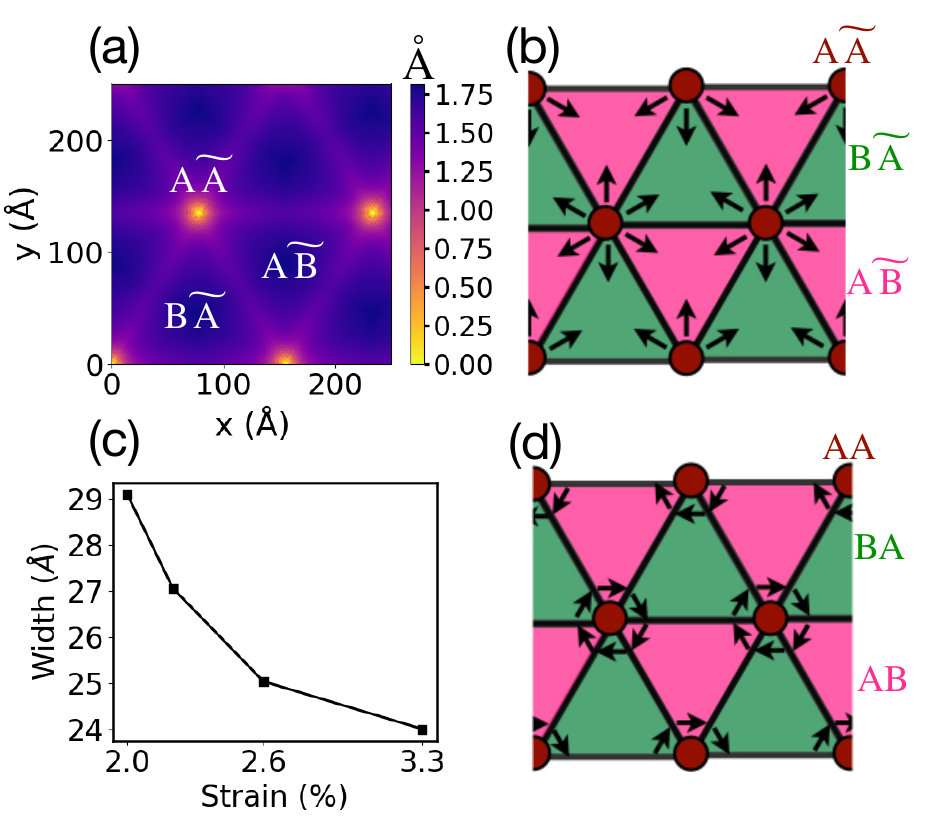}
\caption{\label{fig2} (a): Distribution of order parameter in a MSL induced by 2\%-strain. (b) and (d): The schematic of the solitons in a strain-induced and twist-induced moir\'{e} pattern, respectively. The arrows represent the orientation of order parameters. (c): Variation of domain wall width with applied strain}
\end{figure}

The $A\widetilde{B}$ and $B\widetilde{A}$ stacking regions in the moir\'{e} pattern are separated by domain walls and the $A\widetilde{A}$ stackings which act as point defects.  We characterize the nature of the domain walls and point defects using the order parameter \cite{naik1,maity1,op1,op2}, which is defined as the minimal in-plane displacement vector needed to be applied to the top layer in order to obtain the local bilayer stacking from the highest energy stacking. 
The magnitude of the order parameter is zero at the $A\widetilde{A}$ stacking and maximum at the $A\widetilde{B}$ and
$B\widetilde{A}$ stackings. Fig. \ref{fig2}(a) depicts the distribution of the magnitude of the order parameter in the relaxed moir\'{e} pattern induced by a 2\%-strain. The order parameters point radially outward from the $A\widetilde{A}$ stacking (fig. \ref{fig2}(b)), forming an \textit{aster} defect of topological charge 1. Moreover, the change in
order parameter orientation from $A\widetilde{B}$ stacking region to $B\widetilde{A}$ stacking region, along the longer body diagonal of the supercell, is perpendicular to the domain wall, indicating that they are tensile solitons. We compute the tensile soliton width in biaxially strained MSL for values
of the strain considered (Fig. \ref{fig2}(c)), which increases with the moir\'e length.

The nature of the topological defects and domain walls observed in strain-induced moir\'{e} patterns are different from those in twist-induced moir\'{e} patterns \cite{naik1}. For instance, the order parameter rotates around the $AA$ stacking by 2$\pi$ (Fig. \ref{fig2}(d)), and it is identified as a vortex topological defect \cite{aster1}. Although the order parameter fields appear different in the strain-induced and twist-induced moir\'{e} patterns, they are topologically equivalent. On the other hand, the solitons in the twist-induced moir\'{e} patterns are shear solitons. 

\begin{figure}
\centering
\includegraphics[scale=0.35]{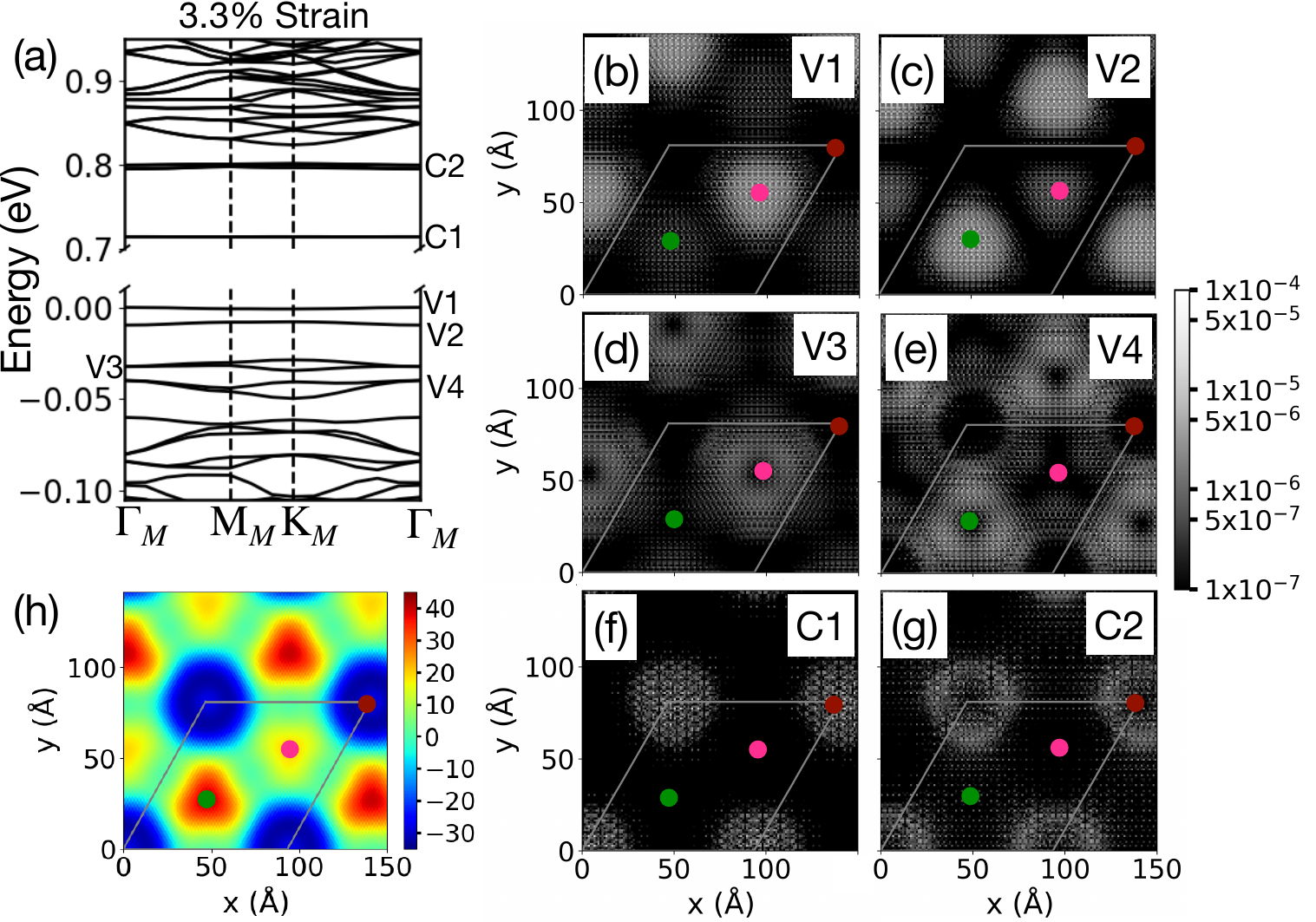}
\caption{\label{fig3} (a): Electronic band structure of a 3.3\% - biaxially strained bilayer MoS$_2$
with the valence band maximum set to zero. (b)-(e): $|\psi_{\Gamma_{M}}(\textbf{r})|^2$
averaged along the out-of-plane direction for the first four distinct energy
levels near the valence band edge. (f)-(g): The same for the first two energy levels at
the conduction band edge. (h): The effective planar potential of the relaxed MSL for the same system. The maroon, pink and green dots represent the $A\tilde{A}$, $A\tilde{B}$ and $B\tilde{A}$ high symmetry regions respectively. The colorbar is in meV. 
}
\end{figure}

\textit{Electronic structure of strain-induced MoS$_2$ MSL with parallel stacking:} Fig. \ref{fig3}(a) shows the
band structure of a 3.3\% - biaxially strained bilayer MoS$_2$ (5223 atoms) along the
$\Gamma_{M}-\textrm{M}_{M}-\textrm{K}_{M}-\Gamma_{M}$ path in the moir\'{e} Brillouin zone (MBZ). Flat bands emerge at both the valence band and conduction band edges. The flat bands at the valence band edge arise from the $\Gamma$ point of the unit cell Brillouin zone (UBZ) \cite{naik1,gammavalley} and possess a trivial topological character \cite{naik1}. The flat bands at the conduction band edge arise from the K point of the UBZ of 
the bottom layer. Due to the external strain applied at the
bottom layer, the K point of the bottom layer moves to a lower energy than the Q or the K point of the top layer. Interestingly, we find that the band gaps of strain-induced MSL 
are significantly smaller compared to those in twisted bilayer MoS$_2$ of similar moir\'{e} length with twist angle
close to 0$^{\circ}$\cite{naik2}. 

Figs.~\ref{fig3}(b)-(e) show that the wavefunctions at the $\Gamma_M$ point($|\psi_{\Gamma_{M}}(\textbf{r})|^2$)  associated with the flat bands near the valence band edge localize in the $A\widetilde{B}$ and $B\widetilde{A}$ regions. While the V1 wavefunction primarily localizes in $A\widetilde{B}$ regions, the V2 localizes in the $B\widetilde{A}$ regions. The V3 and V4, which are doubly degenerate at the $\Gamma$ point also show similar alternate localization patterns. On the other hand, the wavefunctions near the conduction band edge (fig. \ref{fig3}(f)-(g)) localize in the $A\widetilde{A}$ regions of the bottom layer.(See SM for the wave function localization at the K$_M$ point of MBZ.) To gain insight into the localization characteristics, we compute the total Kohn-Sham potential averaged along the out-of-plane direction and filter out the unit-cell-periodic contributions in the in-plane direction.
We keep the first few components of the Fourier expansion of the in-plane potential corresponding to reciprocal lattice vectors of small magnitudes. 
We refer to this filtered out-of-plane averaged potential as the {\em effective planar potential}. Fig.~\ref{fig3}(h) shows the effective planar potential for the 3.3\% strain-induced MSL, which has a three-fold symmetry around the $A\widetilde{A}$. The lack of six-fold symmetry around the $A\widetilde{A}$ stacking is driven by the external strain. The flat valence band wavefunctions localize at the potential hills of the effective potential. 

The asymmetric localization of the wave functions associated with the flat valence bands originates from effective onsite energies which correspond to a staggered potential term in a tight-binding description of the flat bands at the valence band edge. The first two bands can be described by a $s$-orbital tight-binding model on a hexagonal lattice with a staggered potential and the next set of four bands corresponds to a $p_x-p_y$ model with a staggered potential \cite{rubio}.  (See SM, for the details ). Therefore, the biaxial strain-induced MSLs provide a platform for studying the {\em ionic}
Hubbard model \cite{ionichubbard0,ionichubbard1,ionichubbard2,ionichubbard3,ionichubbard4,ionichubbard5,ionichubbard6} on a honeycomb lattice. Motivated by these observations, we estimate the bandwidth ($W$), onsite Coulomb interaction ($U \sim \frac{e^2}{4\pi\varepsilon
a_0}$) and the staggered potential ($\Delta$) associated with the first two bands at the valence band from a $s$-orbital tight-binding  model for the 3.3\% biaxial strain-induced moir\'{e} pattern. We estimate $W$ to be 1.8 meV, $U$ to be  $\sim$ 150 meV ($\varepsilon$ = 3, and $a_0$ is the radius of the localized  distribution), and $\Delta$ is 3.4 meV. All of these parameters are tunable by external strains. Furthermore, the strained MSL 
could be embedded in a high dielectric system or doped sufficiently so that
the effective screening increases to make $U$ comparable to
$\Delta$. This potentially can lead to interesting strongly correlated phenomena~\cite{ionichubbard2,ionichubbard3,ionichubbard4,ionichubbard5,ionichubbard6}. 
Note that the staggered potential is always zero in a twist-induced moir\'{e} pattern due to the absence of asymmetry in the flat-band wavefunctions. 

On the other hand, the flat band wavefunction for the lowest conduction band C1 is localized at the $A\widetilde{A}$ regions and is well separated from the rest of the bands by  $\sim$ 70 meV.
The band is doubly degenerate with a bandwidth $W$ $<$ 1 meV, and its wavefunctions at $\Gamma_{M}$ are strictly localized at $A\widetilde{A}$ with the shape
of a $s$-orbital. Thus, these systems provide an ideal platform for studying Hubbard \cite{hubbard1,hubbard2}
physics on a {\em triangular} lattice. 
We estimate $U$ to be  about 180 meV. The large $\frac{U}{W}$ ratio could potentially lead to the observation of Mott insulating phases at half-filling. 

The electronic structure of MSL with a 2\% external strain shows a qualitatively different behavior, exhibiting the signature of quantum well states (See SM Fig. S1). Several flat bands with bandwidth $<$1 meV emerge near the valence band edge. 
The smaller strain leads to an increased relaxation effect and larger asymmetry between $A\widetilde{B}$ and $B\widetilde{A}$ regions. As a result, the states are localized in either $A\widetilde{B}$ or $B\widetilde{A}$. There are more
flat bands at the conduction band edge for 2\% strain compared to those for 3.3\% strain, and they
localize on $A\widetilde{A}$, resembling states of a triangular
quantum well.

Interestingly, the triangular shape of the $A\widetilde{B}$ and $B\widetilde{A}$ stacking regions can be engineered by applying anisotropic strain resulting into qualitatively similar but quantitatively different localization of the valence band edges (See SM).

\begin{figure}[h]
\centering
\includegraphics[scale=0.3]{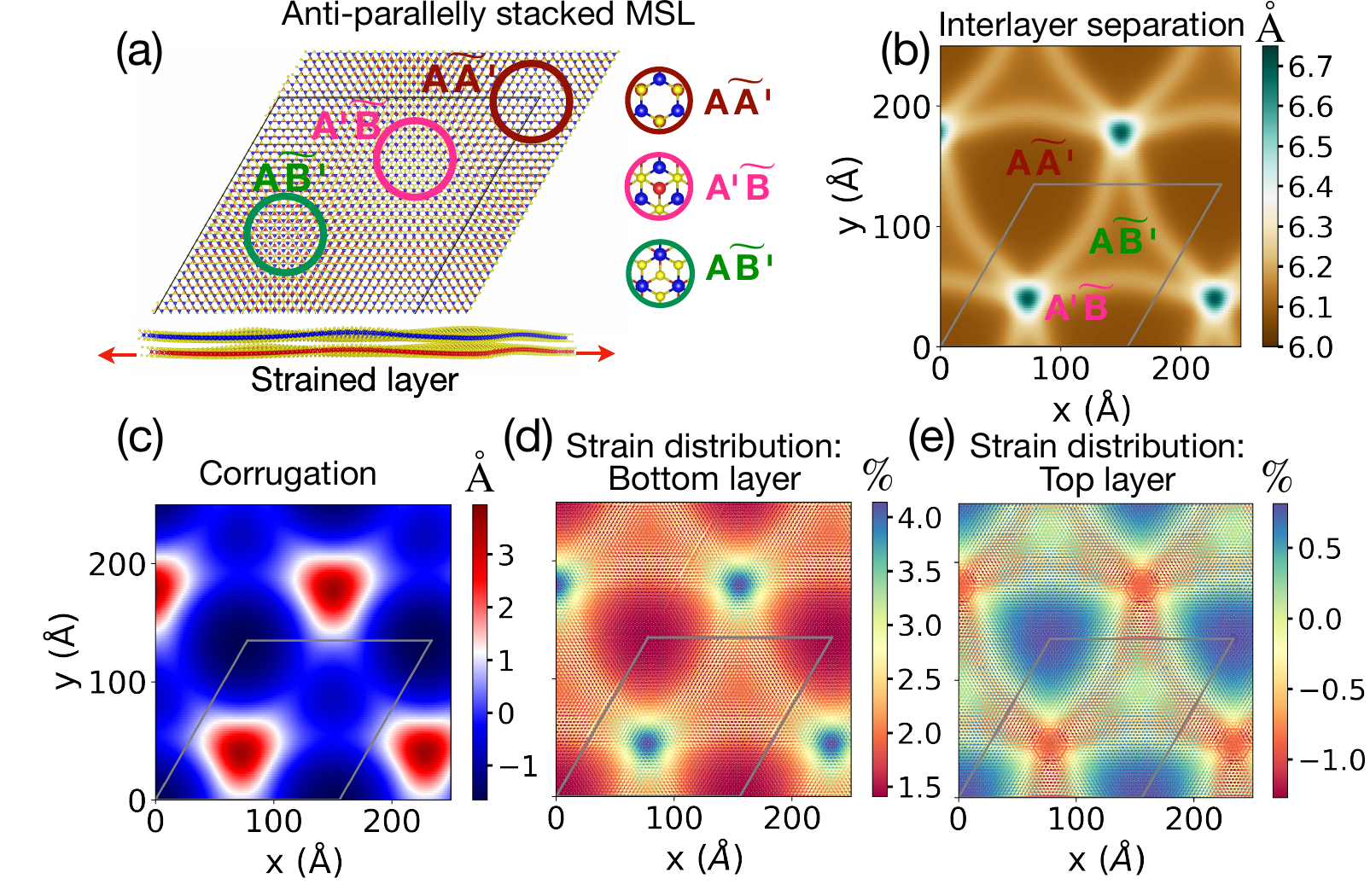}
\caption{\label{fig4} (a): Top and side view of the relaxed atomic structure of a moir\'{e} pattern formed by applying isotropic biaxial strain to the bottom layer of AA$^\prime$ stacked bilayer MoS$_{2}$. The circles denote the high-symmetry stackings. (b) and (c): Interlayer separation and corrugation distribution of a 2\% strain-induced moir\'{e} pattern of MoS$_{2}$.
(d) and (e): Distribution of strains in the bottom and top layers of the same moir\'{e} pattern.}
\end{figure}

\textit{Structural relaxation in strain-induced MoS$_2$ MSL with anti-parallel stacking:} Distinct moir\'{e} patterns compared to those discussed above can be constructed by stretching the bottom layer of an anti-parallelly stacked MoS$_2$ bilayer. Such a moir\'{e} pattern consists of $A\widetilde{A'}$ (M on top of $\widetilde{\textrm{X}}$ and X on top of $\widetilde{\textrm{M}}$), $A\widetilde{B'}$ (Bernal stacking with M on $\widetilde{\textrm{M}}$) and $A'\widetilde{B}$ (Bernal stacking with X on $\widetilde{\textrm{X}}$) stackings (Fig. \ref{fig4}(a)). Conventionally, the high-symmetry stackings of the twisted bilayer with anti-parallel stackings are referred to as AA$^\prime$, AB$^\prime$ and AB$^\prime$, respectively.  

Fig.~\ref{fig4}(a) shows a relaxed biaxially strained moir\'{e} pattern starting from the AA$^\prime$ stacking for which the bottom layer has been stretched equally along the two primitive directions by 2\%. Fig.~\ref{fig4}(b) shows the corresponding interlayer separation landscape. The most stable $A\widetilde{A'}$ stacking regions expand significantly as a result of relaxation and occupy the maximum surface of the moir\'{e} pattern in the shape of Reuleaux triangles. The metastable $A\widetilde{B'}$ stackings have intermediate energy and span a smaller
area. On the other hand, the unfavourable $A'\widetilde{B}$ stackings are confined to the smallest
area. Fig.~\ref{fig4}(c) shows the corresponding corrugation distribution in the relaxed structure. 

Very similar to what happens in the parallelly stacked bilayer MSL, the tensile strain applied to the bottom layer induces a strain in the top layer through the relaxation process. Figs. \ref{fig4}(d) and (e) show the distributions of the strain in the bottom and top layer of the relaxed MSL. 

We can also characterize the domain walls and the point defects using the order parameters. We find that the domain walls are tensile solitons and the point defects are aster topological defects with charge 1.

\begin{figure}
\centering
\includegraphics[scale=0.4]{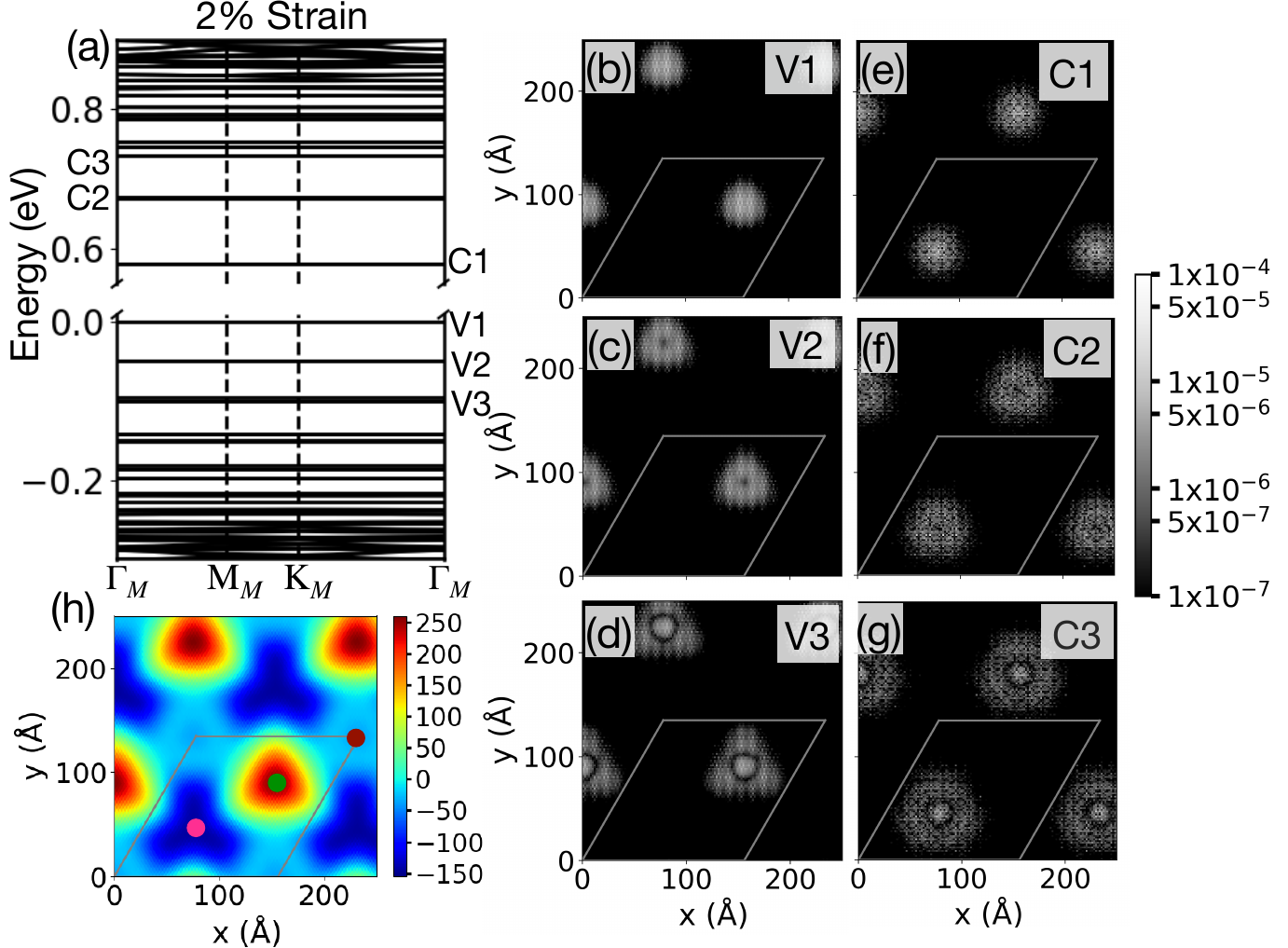}
\caption{\label{fig5} (a): Electronic band structure of 2\% biaxially strained anti-parallely stacked moir\'{e} pattern of MoS$_{2}$ with the valence band maximum set to zero.  (b)-(d) ((e)-(g)): $|\psi_{\Gamma_{M}}(\textbf{r})|^2$ averaged along the out-of-plane direction for the
first three distinct flat bands near the valence band edge (conduction band edge) . (h): The effective planar potential of the relaxed moir\'{e} pattern for the same system. The colorbar is in
meV.}
\end{figure}

\textit{Electronic structure of strain-induced MoS$_2$ MSL with anti-parallel stacking:} Fig.~\ref{fig5}(a) shows the electronic band structure of the 2\% biaxial strain-induced anti-parallely stacked MoS$_{2}$ MSL. We find multiple flat bands both near the valence and the conduction band edges. Specifically, the flat bands up to $\sim$ 100 meV below the valence band edge (V1) and up to $\sim$ 200 meV above the conduction band edge (C1) exhibit bandwidth $<$ 1 meV. Fig.~\ref{fig5}(b)-(g) show the wavefunctions ($|\psi_{\Gamma_{M}}(\textbf{r})|^2$) associated with the first three flat bands near the valence and conduction band edges. The hole wavefunctions for the bands near the valence band edge localize in the $A\widetilde{B'}$ stacking regions and, resemble the solutions of a quantum particle trapped in potential wells. 
The flat bands at the conduction band edge localize in the $A'\widetilde{B}$ stacking regions. 
In order to explain the localization characteristics of the flat band wavefunctions, we show the effective planar potential in Fig. \ref{fig5}(h). The potential has maxima at the $A\widetilde{B'}$ stacking regions and minima at the $A'\widetilde{B}$ stacking regions. The band edges in MSL with anisotropic strain also exhibit similar localization characteristics (See SM).

A spectroscopic experiment on strain-induced MoSe$_2$ MSL with anti-parallel stacking showed that the conduction band edge localizes on $A'\widetilde{B}$ stacking \cite{strainmose2}. Since bilayer MoS$_2$ and MoSe$_2$ exhibit similar electronic structure, our observations apply to strain-induced MoSe$_2$ MSL also, revealing excellent agreement with the experiment.

\textit{Comparison of structural and electronic properties: twist vs. strain:} While the structural relaxation and electronic properties of the moir\'{e} pattern formed by biaxial strains have some similarities to those of the moir\'{e} pattern formed by twist, there are many important differences, as well. We summarize the differences between the two types of MSLs in Table 1. In particular, we note that while the relaxed structures look similar, the electronic structure are qualitatively different indicating the importance of such \textit{ab initio} studies.
\begin{table}[]
\caption{Comparison of properties between strained and twisted bilayer TMDs}
\begin{tabular}{c|c|c}
                    & \textbf{Strained}                                                            & \textbf{Twisted}                                                            \\ \hline
\multicolumn{3}{l}{\textbf{Structure}}                                                                                                                                  \\ \hline
Point defect        & Aster                                                               & Vortex                                                             \\ \hline
Soliton             & Tensile                                                             & Shear                                                              \\ \hline
\multicolumn{3}{l}{\textbf{Electronic structure}}                                                                                                                       \\ \hline
\begin{tabular}[c]{@{}c@{}}VBM states for\\parallel stacking \end{tabular}  & \begin{tabular}[c]{@{}c@{}}Asymmetrically\\localized on\\ $A\widetilde{B}$ and $B\widetilde{A}$\end{tabular} & \begin{tabular}[c]{@{}c@{}}Symmetrically\\ localized on\\ AB and BA\end{tabular} \\ \hline
\begin{tabular}[c]{@{}c@{}}CBM states for\\parallel stacking \end{tabular}    & $A\widetilde{A}$                                                                  & Delocalized                                                        \\ \hline
\begin{tabular}[c]{@{}c@{}}VBM states for\\antiparallel stacking \end{tabular} & $A\widetilde{B}'$                                                                 & AA'                                                                \\ \hline
\begin{tabular}[c]{@{}c@{}}CBM states for\\antiparallel stacking \end{tabular} & $A'\widetilde{B}$                                                                 & AB'                                                               
\end{tabular}
\end{table}

\textit{Conclusion} We have studied the properties of strain-induced TMD MSLs and compared those to of twist-induced TMD MSLs.
The strain-induced MSLs differ from twist-induced MSLs in
both structural and electronic properties.
Although the relaxation patterns are similar,
the order parameter fields are different for the two types of MSLs; the order parameter
field forms aster defects in strain-induced MSLs, in contrast to vortex defects in
twist-induced MSLs.  But as they are
topologically equivalent, with the same winding number, the transition from aster to vortex defect can take place smoothly
through the formation of a spiral defect, and it will be worthwhile to study
strain-induced MSLs with a non-zero twist angle. 
In addition, due to the large strain accumulated at the aster defect sites in strain-induced MSLs, we expect that it can be perturbed relatively easily by the inhomogeneous strain invariably present in devices, while the vortex defect is expected to be more robust. 
Both the MSLs with parallel and anti-parallel stackings host flat bands, and our findings on their localization characteristics are consistent with experimental observations.
The flat bands are well
separated and ideal for studying strongly correlated phenomena in the
Hubbard model and the ionic Hubbard model. The extent of localization of the wave functions and the
separation between their centers which determine the
onsite Coulomb interaction and the hopping terms, respectively, can be tuned by the applied strain. Thus strain-induced
MSLs provide a promising parallel pathway for investigating and manipulating hole and electron
properties in TMD MSLs.

We acknowledge the Supercomputer and Education Resource Center at the Indian Institute
of Science for providing computational facilities. M. J. gratefully acknowledges
financial support through grant no. DST/NSM/R\&D\_HPC\_Applications/2021/23 from the
National Supercomputing Mission of the Department of Science and Technology, and No. DST/NM/TUE/QM-10/2019 through the Nano Mission of the Department of Science and Technology, India.
H. R. K gratefully acknowledges financial support from the  Indian National Science Academy under grant number No. INSA/SP/SS/2023/, and through grant no. SB/DF/005/2017
from the Science and Engineering Research Board of the Department of Science
and Technology, India.

%

\begin{thebibliography}{59}%
\makeatletter
\providecommand \@ifxundefined [1]{%
 \@ifx{#1\undefined}
}%
\providecommand \@ifnum [1]{%
 \ifnum #1\expandafter \@firstoftwo
 \else \expandafter \@secondoftwo
 \fi
}%
\providecommand \@ifx [1]{%
 \ifx #1\expandafter \@firstoftwo
 \else \expandafter \@secondoftwo
 \fi
}%
\providecommand \natexlab [1]{#1}%
\providecommand \enquote  [1]{``#1''}%
\providecommand \bibnamefont  [1]{#1}%
\providecommand \bibfnamefont [1]{#1}%
\providecommand \citenamefont [1]{#1}%
\providecommand \href@noop [0]{\@secondoftwo}%
\providecommand \href [0]{\begingroup \@sanitize@url \@href}%
\providecommand \@href[1]{\@@startlink{#1}\@@href}%
\providecommand \@@href[1]{\endgroup#1\@@endlink}%
\providecommand \@sanitize@url [0]{\catcode `\\12\catcode `\$12\catcode
  `\&12\catcode `\#12\catcode `\^12\catcode `\_12\catcode `\%12\relax}%
\providecommand \@@startlink[1]{}%
\providecommand \@@endlink[0]{}%
\providecommand \url  [0]{\begingroup\@sanitize@url \@url }%
\providecommand \@url [1]{\endgroup\@href {#1}{\urlprefix }}%
\providecommand \urlprefix  [0]{URL }%
\providecommand \Eprint [0]{\href }%
\providecommand \doibase [0]{https://doi.org/}%
\providecommand \selectlanguage [0]{\@gobble}%
\providecommand \bibinfo  [0]{\@secondoftwo}%
\providecommand \bibfield  [0]{\@secondoftwo}%
\providecommand \translation [1]{[#1]}%
\providecommand \BibitemOpen [0]{}%
\providecommand \bibitemStop [0]{}%
\providecommand \bibitemNoStop [0]{.\EOS\space}%
\providecommand \EOS [0]{\spacefactor3000\relax}%
\providecommand \BibitemShut  [1]{\csname bibitem#1\endcsname}%
\let\auto@bib@innerbib\@empty
\bibitem [{\citenamefont {Edelberg}\ \emph {et~al.}(2020)\citenamefont
  {Edelberg}, \citenamefont {Kumar}, \citenamefont {Shenoy}, \citenamefont
  {Ochoa},\ and\ \citenamefont {Pasupathy}}]{strainmose2}%
  \BibitemOpen
  \bibfield  {author} {\bibinfo {author} {\bibfnamefont {D.}~\bibnamefont
  {Edelberg}}, \bibinfo {author} {\bibfnamefont {H.}~\bibnamefont {Kumar}},
  \bibinfo {author} {\bibfnamefont {V.}~\bibnamefont {Shenoy}}, \bibinfo
  {author} {\bibfnamefont {H.}~\bibnamefont {Ochoa}},\ and\ \bibinfo {author}
  {\bibfnamefont {A.~N.}\ \bibnamefont {Pasupathy}},\ }\bibfield  {title}
  {\bibinfo {title} {Tunable strain soliton networks confine electrons in van
  der waals materials},\ }\href@noop {} {\bibfield  {journal} {\bibinfo
  {journal} {Nature Physics}\ }\textbf {\bibinfo {volume} {16}},\ \bibinfo
  {pages} {1097} (\bibinfo {year} {2020})}\BibitemShut {NoStop}%
\bibitem [{\citenamefont {Cao}\ \emph {et~al.}(2018{\natexlab{a}})\citenamefont
  {Cao}, \citenamefont {Fatemi}, \citenamefont {Demir}, \citenamefont {Fang},
  \citenamefont {Tomarken}, \citenamefont {Luo}, \citenamefont
  {Sanchez-Yamagishi}, \citenamefont {Watanabe}, \citenamefont {Taniguchi},
  \citenamefont {Kaxiras}, \citenamefont {Ashoori},\ and\ \citenamefont
  {Jarillo-Herrero}}]{graphene2}%
  \BibitemOpen
  \bibfield  {author} {\bibinfo {author} {\bibfnamefont {Y.}~\bibnamefont
  {Cao}}, \bibinfo {author} {\bibfnamefont {V.}~\bibnamefont {Fatemi}},
  \bibinfo {author} {\bibfnamefont {A.}~\bibnamefont {Demir}}, \bibinfo
  {author} {\bibfnamefont {S.}~\bibnamefont {Fang}}, \bibinfo {author}
  {\bibfnamefont {S.~L.}\ \bibnamefont {Tomarken}}, \bibinfo {author}
  {\bibfnamefont {J.~Y.}\ \bibnamefont {Luo}}, \bibinfo {author} {\bibfnamefont
  {J.~D.}\ \bibnamefont {Sanchez-Yamagishi}}, \bibinfo {author} {\bibfnamefont
  {K.}~\bibnamefont {Watanabe}}, \bibinfo {author} {\bibfnamefont
  {T.}~\bibnamefont {Taniguchi}}, \bibinfo {author} {\bibfnamefont
  {E.}~\bibnamefont {Kaxiras}}, \bibinfo {author} {\bibfnamefont {R.~C.}\
  \bibnamefont {Ashoori}},\ and\ \bibinfo {author} {\bibfnamefont
  {P.}~\bibnamefont {Jarillo-Herrero}},\ }\bibfield  {title} {\bibinfo {title}
  {Correlated insulator behaviour at half-filling in magic-angle graphene
  superlattices},\ }\href@noop {} {\bibfield  {journal} {\bibinfo  {journal}
  {Nature}\ }\textbf {\bibinfo {volume} {556}},\ \bibinfo {pages} {80}
  (\bibinfo {year} {2018}{\natexlab{a}})}\BibitemShut {NoStop}%
\bibitem [{\citenamefont {Zhou}\ \emph {et~al.}(2021)\citenamefont {Zhou},
  \citenamefont {Sung}, \citenamefont {Brutschea}, \citenamefont {Esterlis},
  \citenamefont {Wang}, \citenamefont {Scuri}, \citenamefont {Gelly},
  \citenamefont {Heo}, \citenamefont {Taniguchi}, \citenamefont {Watanabe}
  \emph {et~al.}}]{expt2}%
  \BibitemOpen
  \bibfield  {author} {\bibinfo {author} {\bibfnamefont {Y.}~\bibnamefont
  {Zhou}}, \bibinfo {author} {\bibfnamefont {J.}~\bibnamefont {Sung}}, \bibinfo
  {author} {\bibfnamefont {E.}~\bibnamefont {Brutschea}}, \bibinfo {author}
  {\bibfnamefont {I.}~\bibnamefont {Esterlis}}, \bibinfo {author}
  {\bibfnamefont {Y.}~\bibnamefont {Wang}}, \bibinfo {author} {\bibfnamefont
  {G.}~\bibnamefont {Scuri}}, \bibinfo {author} {\bibfnamefont {R.~J.}\
  \bibnamefont {Gelly}}, \bibinfo {author} {\bibfnamefont {H.}~\bibnamefont
  {Heo}}, \bibinfo {author} {\bibfnamefont {T.}~\bibnamefont {Taniguchi}},
  \bibinfo {author} {\bibfnamefont {K.}~\bibnamefont {Watanabe}}, \emph
  {et~al.},\ }\bibfield  {title} {\bibinfo {title} {Bilayer wigner crystals in
  a transition metal dichalcogenide heterostructure},\ }\href@noop {}
  {\bibfield  {journal} {\bibinfo  {journal} {Nature}\ }\textbf {\bibinfo
  {volume} {595}},\ \bibinfo {pages} {48} (\bibinfo {year} {2021})}\BibitemShut
  {NoStop}%
\bibitem [{\citenamefont {Wang}\ \emph {et~al.}(2020)\citenamefont {Wang},
  \citenamefont {Shih}, \citenamefont {Ghiotto}, \citenamefont {Xian},
  \citenamefont {Rhodes}, \citenamefont {Tan}, \citenamefont {Claassen},
  \citenamefont {Kennes}, \citenamefont {Bai}, \citenamefont {Kim},
  \citenamefont {Watanabe}, \citenamefont {Taniguchi}, \citenamefont {Zhu},
  \citenamefont {Hone}, \citenamefont {Rubio}, \citenamefont {Pasupathy},\ and\
  \citenamefont {Dean}}]{wse2_expt2}%
  \BibitemOpen
  \bibfield  {author} {\bibinfo {author} {\bibfnamefont {L.}~\bibnamefont
  {Wang}}, \bibinfo {author} {\bibfnamefont {E.-M.}\ \bibnamefont {Shih}},
  \bibinfo {author} {\bibfnamefont {A.}~\bibnamefont {Ghiotto}}, \bibinfo
  {author} {\bibfnamefont {L.}~\bibnamefont {Xian}}, \bibinfo {author}
  {\bibfnamefont {D.~A.}\ \bibnamefont {Rhodes}}, \bibinfo {author}
  {\bibfnamefont {C.}~\bibnamefont {Tan}}, \bibinfo {author} {\bibfnamefont
  {M.}~\bibnamefont {Claassen}}, \bibinfo {author} {\bibfnamefont {D.~M.}\
  \bibnamefont {Kennes}}, \bibinfo {author} {\bibfnamefont {Y.}~\bibnamefont
  {Bai}}, \bibinfo {author} {\bibfnamefont {B.}~\bibnamefont {Kim}}, \bibinfo
  {author} {\bibfnamefont {K.}~\bibnamefont {Watanabe}}, \bibinfo {author}
  {\bibfnamefont {T.}~\bibnamefont {Taniguchi}}, \bibinfo {author}
  {\bibfnamefont {X.}~\bibnamefont {Zhu}}, \bibinfo {author} {\bibfnamefont
  {J.}~\bibnamefont {Hone}}, \bibinfo {author} {\bibfnamefont {A.}~\bibnamefont
  {Rubio}}, \bibinfo {author} {\bibfnamefont {A.~N.}\ \bibnamefont
  {Pasupathy}},\ and\ \bibinfo {author} {\bibfnamefont {C.~R.}\ \bibnamefont
  {Dean}},\ }\bibfield  {title} {\bibinfo {title} {Correlated electronic phases
  in twisted bilayer transition metal dichalcogenides},\ }\href@noop {}
  {\bibfield  {journal} {\bibinfo  {journal} {Nature Materials}\ }\textbf
  {\bibinfo {volume} {19}},\ \bibinfo {pages} {861} (\bibinfo {year}
  {2020})}\BibitemShut {NoStop}%
\bibitem [{\citenamefont {Cao}\ \emph {et~al.}(2018{\natexlab{b}})\citenamefont
  {Cao}, \citenamefont {Fatemi}, \citenamefont {Fang}, \citenamefont
  {Watanabe}, \citenamefont {Taniguchi}, \citenamefont {Kaxiras},\ and\
  \citenamefont {Jarillo-Herrero}}]{graphene1}%
  \BibitemOpen
  \bibfield  {author} {\bibinfo {author} {\bibfnamefont {Y.}~\bibnamefont
  {Cao}}, \bibinfo {author} {\bibfnamefont {V.}~\bibnamefont {Fatemi}},
  \bibinfo {author} {\bibfnamefont {S.}~\bibnamefont {Fang}}, \bibinfo {author}
  {\bibfnamefont {K.}~\bibnamefont {Watanabe}}, \bibinfo {author}
  {\bibfnamefont {T.}~\bibnamefont {Taniguchi}}, \bibinfo {author}
  {\bibfnamefont {E.}~\bibnamefont {Kaxiras}},\ and\ \bibinfo {author}
  {\bibfnamefont {P.}~\bibnamefont {Jarillo-Herrero}},\ }\bibfield  {title}
  {\bibinfo {title} {Unconventional superconductivity in magic-angle graphene
  superlattices},\ }\href@noop {} {\bibfield  {journal} {\bibinfo  {journal}
  {Nature}\ }\textbf {\bibinfo {volume} {556}},\ \bibinfo {pages} {43}
  (\bibinfo {year} {2018}{\natexlab{b}})}\BibitemShut {NoStop}%
\bibitem [{\citenamefont {Liu}\ \emph {et~al.}(2021)\citenamefont {Liu},
  \citenamefont {Xian}, \citenamefont {Mu}, \citenamefont {Zhao}, \citenamefont
  {Liu}, \citenamefont {Rubio},\ and\ \citenamefont {Wang}}]{rubio3}%
  \BibitemOpen
  \bibfield  {author} {\bibinfo {author} {\bibfnamefont {B.}~\bibnamefont
  {Liu}}, \bibinfo {author} {\bibfnamefont {L.}~\bibnamefont {Xian}}, \bibinfo
  {author} {\bibfnamefont {H.}~\bibnamefont {Mu}}, \bibinfo {author}
  {\bibfnamefont {G.}~\bibnamefont {Zhao}}, \bibinfo {author} {\bibfnamefont
  {Z.}~\bibnamefont {Liu}}, \bibinfo {author} {\bibfnamefont {A.}~\bibnamefont
  {Rubio}},\ and\ \bibinfo {author} {\bibfnamefont {Z.~F.}\ \bibnamefont
  {Wang}},\ }\bibfield  {title} {\bibinfo {title} {Higher-order band topology
  in twisted moir\'e superlattice},\ }\href@noop {} {\bibfield  {journal}
  {\bibinfo  {journal} {Phys. Rev. Lett.}\ }\textbf {\bibinfo {volume} {126}},\
  \bibinfo {pages} {066401} (\bibinfo {year} {2021})}\BibitemShut {NoStop}%
\bibitem [{\citenamefont {Vogl}\ \emph {et~al.}(2021)\citenamefont {Vogl},
  \citenamefont {Rodriguez-Vega}, \citenamefont {Flebus}, \citenamefont
  {MacDonald},\ and\ \citenamefont {Fiete}}]{macdonald3}%
  \BibitemOpen
  \bibfield  {author} {\bibinfo {author} {\bibfnamefont {M.}~\bibnamefont
  {Vogl}}, \bibinfo {author} {\bibfnamefont {M.}~\bibnamefont
  {Rodriguez-Vega}}, \bibinfo {author} {\bibfnamefont {B.}~\bibnamefont
  {Flebus}}, \bibinfo {author} {\bibfnamefont {A.~H.}\ \bibnamefont
  {MacDonald}},\ and\ \bibinfo {author} {\bibfnamefont {G.~A.}\ \bibnamefont
  {Fiete}},\ }\bibfield  {title} {\bibinfo {title} {Floquet engineering of
  topological transitions in a twisted transition metal dichalcogenide
  homobilayer},\ }\href@noop {} {\bibfield  {journal} {\bibinfo  {journal}
  {Phys. Rev. B}\ }\textbf {\bibinfo {volume} {103}},\ \bibinfo {pages}
  {014310} (\bibinfo {year} {2021})}\BibitemShut {NoStop}%
\bibitem [{\citenamefont {Zeng}\ \emph {et~al.}(2023)\citenamefont {Zeng},
  \citenamefont {Xia}, \citenamefont {Kang}, \citenamefont {Zhu}, \citenamefont
  {Kn{\"u}ppel}, \citenamefont {Vaswani}, \citenamefont {Watanabe},
  \citenamefont {Taniguchi}, \citenamefont {Mak},\ and\ \citenamefont
  {Shan}}]{zeng2023thermodynamic}%
  \BibitemOpen
  \bibfield  {author} {\bibinfo {author} {\bibfnamefont {Y.}~\bibnamefont
  {Zeng}}, \bibinfo {author} {\bibfnamefont {Z.}~\bibnamefont {Xia}}, \bibinfo
  {author} {\bibfnamefont {K.}~\bibnamefont {Kang}}, \bibinfo {author}
  {\bibfnamefont {J.}~\bibnamefont {Zhu}}, \bibinfo {author} {\bibfnamefont
  {P.}~\bibnamefont {Kn{\"u}ppel}}, \bibinfo {author} {\bibfnamefont
  {C.}~\bibnamefont {Vaswani}}, \bibinfo {author} {\bibfnamefont
  {K.}~\bibnamefont {Watanabe}}, \bibinfo {author} {\bibfnamefont
  {T.}~\bibnamefont {Taniguchi}}, \bibinfo {author} {\bibfnamefont {K.~F.}\
  \bibnamefont {Mak}},\ and\ \bibinfo {author} {\bibfnamefont {J.}~\bibnamefont
  {Shan}},\ }\bibfield  {title} {\bibinfo {title} {Thermodynamic evidence of
  fractional chern insulator in moir{\'e} mote2},\ }\href@noop {} {\bibfield
  {journal} {\bibinfo  {journal} {Nature}\ }\textbf {\bibinfo {volume} {622}},\
  \bibinfo {pages} {69} (\bibinfo {year} {2023})}\BibitemShut {NoStop}%
\bibitem [{\citenamefont {Qi}\ \emph {et~al.}(2020)\citenamefont {Qi},
  \citenamefont {Fu}, \citenamefont {Sun},\ and\ \citenamefont {Gu}}]{fu4}%
  \BibitemOpen
  \bibfield  {author} {\bibinfo {author} {\bibfnamefont {Y.}~\bibnamefont
  {Qi}}, \bibinfo {author} {\bibfnamefont {L.}~\bibnamefont {Fu}}, \bibinfo
  {author} {\bibfnamefont {K.}~\bibnamefont {Sun}},\ and\ \bibinfo {author}
  {\bibfnamefont {Z.}~\bibnamefont {Gu}},\ }\bibfield  {title} {\bibinfo
  {title} {Coexistence of antiferromagnetism and topological superconductivity
  on the honeycomb lattice hubbard model},\ }\href@noop {} {\bibfield
  {journal} {\bibinfo  {journal} {Phys. Rev. B}\ }\textbf {\bibinfo {volume}
  {102}},\ \bibinfo {pages} {245140} (\bibinfo {year} {2020})}\BibitemShut
  {NoStop}%
\bibitem [{\citenamefont {Bai}\ \emph {et~al.}(2020)\citenamefont {Bai},
  \citenamefont {Zhou}, \citenamefont {Wang}, \citenamefont {Wu}, \citenamefont
  {McGilly}, \citenamefont {Halbertal}, \citenamefont {Lo}, \citenamefont
  {Liu}, \citenamefont {Ardelean}, \citenamefont {Rivera}, \citenamefont
  {Finney}, \citenamefont {Yang}, \citenamefont {Basov}, \citenamefont {Yao},
  \citenamefont {Xu}, \citenamefont {Hone}, \citenamefont {Pasupathy},\ and\
  \citenamefont {Zhu}}]{expt1}%
  \BibitemOpen
  \bibfield  {author} {\bibinfo {author} {\bibfnamefont {Y.}~\bibnamefont
  {Bai}}, \bibinfo {author} {\bibfnamefont {L.}~\bibnamefont {Zhou}}, \bibinfo
  {author} {\bibfnamefont {J.}~\bibnamefont {Wang}}, \bibinfo {author}
  {\bibfnamefont {W.}~\bibnamefont {Wu}}, \bibinfo {author} {\bibfnamefont
  {L.~J.}\ \bibnamefont {McGilly}}, \bibinfo {author} {\bibfnamefont
  {D.}~\bibnamefont {Halbertal}}, \bibinfo {author} {\bibfnamefont {C.~F.~B.}\
  \bibnamefont {Lo}}, \bibinfo {author} {\bibfnamefont {F.}~\bibnamefont
  {Liu}}, \bibinfo {author} {\bibfnamefont {J.}~\bibnamefont {Ardelean}},
  \bibinfo {author} {\bibfnamefont {P.}~\bibnamefont {Rivera}}, \bibinfo
  {author} {\bibfnamefont {N.~R.}\ \bibnamefont {Finney}}, \bibinfo {author}
  {\bibfnamefont {X.-C.}\ \bibnamefont {Yang}}, \bibinfo {author}
  {\bibfnamefont {D.~N.}\ \bibnamefont {Basov}}, \bibinfo {author}
  {\bibfnamefont {W.}~\bibnamefont {Yao}}, \bibinfo {author} {\bibfnamefont
  {X.}~\bibnamefont {Xu}}, \bibinfo {author} {\bibfnamefont {J.}~\bibnamefont
  {Hone}}, \bibinfo {author} {\bibfnamefont {A.~N.}\ \bibnamefont
  {Pasupathy}},\ and\ \bibinfo {author} {\bibfnamefont {X.~Y.}\ \bibnamefont
  {Zhu}},\ }\bibfield  {title} {\bibinfo {title} {Excitons in strain-induced
  one-dimensional moir{\'e}potentials at transition metal dichalcogenide
  heterojunctions},\ }\href@noop {} {\bibfield  {journal} {\bibinfo  {journal}
  {Nature Materials}\ }\textbf {\bibinfo {volume} {19}},\ \bibinfo {pages}
  {1068} (\bibinfo {year} {2020})}\BibitemShut {NoStop}%
\bibitem [{\citenamefont {Sung}\ \emph {et~al.}(2020)\citenamefont {Sung},
  \citenamefont {Zhou}, \citenamefont {Scuri}, \citenamefont {Z{\'o}lyomi},
  \citenamefont {Andersen}, \citenamefont {Yoo}, \citenamefont {Wild},
  \citenamefont {Joe}, \citenamefont {Gelly}, \citenamefont {Heo},
  \citenamefont {Magorrian}, \citenamefont {B{\'e}rub{\'e}}, \citenamefont
  {Valdivia}, \citenamefont {Taniguchi}, \citenamefont {Watanabe},
  \citenamefont {Lukin}, \citenamefont {Kim}, \citenamefont {Fal'ko},\ and\
  \citenamefont {Park}}]{expt3}%
  \BibitemOpen
  \bibfield  {author} {\bibinfo {author} {\bibfnamefont {J.}~\bibnamefont
  {Sung}}, \bibinfo {author} {\bibfnamefont {Y.}~\bibnamefont {Zhou}}, \bibinfo
  {author} {\bibfnamefont {G.}~\bibnamefont {Scuri}}, \bibinfo {author}
  {\bibfnamefont {V.}~\bibnamefont {Z{\'o}lyomi}}, \bibinfo {author}
  {\bibfnamefont {T.~I.}\ \bibnamefont {Andersen}}, \bibinfo {author}
  {\bibfnamefont {H.}~\bibnamefont {Yoo}}, \bibinfo {author} {\bibfnamefont
  {D.~S.}\ \bibnamefont {Wild}}, \bibinfo {author} {\bibfnamefont {A.~Y.}\
  \bibnamefont {Joe}}, \bibinfo {author} {\bibfnamefont {R.~J.}\ \bibnamefont
  {Gelly}}, \bibinfo {author} {\bibfnamefont {H.}~\bibnamefont {Heo}}, \bibinfo
  {author} {\bibfnamefont {S.~J.}\ \bibnamefont {Magorrian}}, \bibinfo {author}
  {\bibfnamefont {D.}~\bibnamefont {B{\'e}rub{\'e}}}, \bibinfo {author}
  {\bibfnamefont {A.~M.~M.}\ \bibnamefont {Valdivia}}, \bibinfo {author}
  {\bibfnamefont {T.}~\bibnamefont {Taniguchi}}, \bibinfo {author}
  {\bibfnamefont {K.}~\bibnamefont {Watanabe}}, \bibinfo {author}
  {\bibfnamefont {M.~D.}\ \bibnamefont {Lukin}}, \bibinfo {author}
  {\bibfnamefont {P.}~\bibnamefont {Kim}}, \bibinfo {author} {\bibfnamefont
  {V.~I.}\ \bibnamefont {Fal'ko}},\ and\ \bibinfo {author} {\bibfnamefont
  {H.}~\bibnamefont {Park}},\ }\bibfield  {title} {\bibinfo {title} {Broken
  mirror symmetry in excitonic response of reconstructed domains in twisted
  mose2/mose2 bilayers},\ }\href {https://doi.org/10.1038/s41565-020-0728-z}
  {\bibfield  {journal} {\bibinfo  {journal} {Nature Nanotechnology}\ }\textbf
  {\bibinfo {volume} {15}},\ \bibinfo {pages} {750} (\bibinfo {year}
  {2020})}\BibitemShut {NoStop}%
\bibitem [{\citenamefont {Forg}\ \emph {et~al.}(2021)\citenamefont {Forg},
  \citenamefont {Baimuratov}, \citenamefont {Kruchinin}, \citenamefont {Vovk},
  \citenamefont {Scherzer}, \citenamefont {Forste}, \citenamefont {Funk},
  \citenamefont {Watanabe}, \citenamefont {Taniguchi},\ and\ \citenamefont
  {Hogele}}]{expt4}%
  \BibitemOpen
  \bibfield  {author} {\bibinfo {author} {\bibfnamefont {M.}~\bibnamefont
  {Forg}}, \bibinfo {author} {\bibfnamefont {A.~S.}\ \bibnamefont
  {Baimuratov}}, \bibinfo {author} {\bibfnamefont {S.~Y.}\ \bibnamefont
  {Kruchinin}}, \bibinfo {author} {\bibfnamefont {I.~A.}\ \bibnamefont {Vovk}},
  \bibinfo {author} {\bibfnamefont {J.}~\bibnamefont {Scherzer}}, \bibinfo
  {author} {\bibfnamefont {J.}~\bibnamefont {Forste}}, \bibinfo {author}
  {\bibfnamefont {V.}~\bibnamefont {Funk}}, \bibinfo {author} {\bibfnamefont
  {K.}~\bibnamefont {Watanabe}}, \bibinfo {author} {\bibfnamefont
  {T.}~\bibnamefont {Taniguchi}},\ and\ \bibinfo {author} {\bibfnamefont
  {A.}~\bibnamefont {Hogele}},\ }\bibfield  {title} {\bibinfo {title} {Moir\'e
  excitons in mose2-wse2 heterobilayers and heterotrilayers},\ }\href@noop {}
  {\bibfield  {journal} {\bibinfo  {journal} {Nature Communications}\ }\textbf
  {\bibinfo {volume} {12}},\ \bibinfo {pages} {1656} (\bibinfo {year}
  {2021})}\BibitemShut {NoStop}%
\bibitem [{\citenamefont {Li}\ \emph {et~al.}(2021)\citenamefont {Li},
  \citenamefont {Lu}, \citenamefont {Leon}, \citenamefont {Hou}, \citenamefont
  {Lu}, \citenamefont {Kaczmarek}, \citenamefont {Lyu}, \citenamefont
  {Taniguchi}, \citenamefont {Watanabe}, \citenamefont {Zhao}, \citenamefont
  {Yang},\ and\ \citenamefont {Deotare}}]{expt5}%
  \BibitemOpen
  \bibfield  {author} {\bibinfo {author} {\bibfnamefont {Z.}~\bibnamefont
  {Li}}, \bibinfo {author} {\bibfnamefont {X.}~\bibnamefont {Lu}}, \bibinfo
  {author} {\bibfnamefont {D.~F.~C.}\ \bibnamefont {Leon}}, \bibinfo {author}
  {\bibfnamefont {J.}~\bibnamefont {Hou}}, \bibinfo {author} {\bibfnamefont
  {Y.}~\bibnamefont {Lu}}, \bibinfo {author} {\bibfnamefont {A.}~\bibnamefont
  {Kaczmarek}}, \bibinfo {author} {\bibfnamefont {Z.}~\bibnamefont {Lyu}},
  \bibinfo {author} {\bibfnamefont {T.}~\bibnamefont {Taniguchi}}, \bibinfo
  {author} {\bibfnamefont {K.}~\bibnamefont {Watanabe}}, \bibinfo {author}
  {\bibfnamefont {L.}~\bibnamefont {Zhao}}, \bibinfo {author} {\bibfnamefont
  {L.}~\bibnamefont {Yang}},\ and\ \bibinfo {author} {\bibfnamefont {P.~B.}\
  \bibnamefont {Deotare}},\ }\bibfield  {title} {\bibinfo {title} {Exciton
  transport under periodic potential in mose2/wse2 heterostructures},\
  }\href@noop {} {\bibfield  {journal} {\bibinfo  {journal} {ACS Nano}\
  }\textbf {\bibinfo {volume} {15}},\ \bibinfo {pages} {1539} (\bibinfo {year}
  {2021})}\BibitemShut {NoStop}%
\bibitem [{\citenamefont {Andersen}\ \emph {et~al.}(2021)\citenamefont
  {Andersen}, \citenamefont {Scuri}, \citenamefont {Sushko}, \citenamefont
  {De~Greve}, \citenamefont {Sung}, \citenamefont {Zhou}, \citenamefont {Wild},
  \citenamefont {Gelly}, \citenamefont {Heo}, \citenamefont {B{\'e}rub{\'e}},
  \citenamefont {Joe}, \citenamefont {Jauregui}, \citenamefont {Watanabe},
  \citenamefont {Taniguchi}, \citenamefont {Kim}, \citenamefont {Park},\ and\
  \citenamefont {Lukin}}]{expt6}%
  \BibitemOpen
  \bibfield  {author} {\bibinfo {author} {\bibfnamefont {T.~I.}\ \bibnamefont
  {Andersen}}, \bibinfo {author} {\bibfnamefont {G.}~\bibnamefont {Scuri}},
  \bibinfo {author} {\bibfnamefont {A.}~\bibnamefont {Sushko}}, \bibinfo
  {author} {\bibfnamefont {K.}~\bibnamefont {De~Greve}}, \bibinfo {author}
  {\bibfnamefont {J.}~\bibnamefont {Sung}}, \bibinfo {author} {\bibfnamefont
  {Y.}~\bibnamefont {Zhou}}, \bibinfo {author} {\bibfnamefont {D.~S.}\
  \bibnamefont {Wild}}, \bibinfo {author} {\bibfnamefont {R.~J.}\ \bibnamefont
  {Gelly}}, \bibinfo {author} {\bibfnamefont {H.}~\bibnamefont {Heo}}, \bibinfo
  {author} {\bibfnamefont {D.}~\bibnamefont {B{\'e}rub{\'e}}}, \bibinfo
  {author} {\bibfnamefont {A.~Y.}\ \bibnamefont {Joe}}, \bibinfo {author}
  {\bibfnamefont {L.~A.}\ \bibnamefont {Jauregui}}, \bibinfo {author}
  {\bibfnamefont {K.}~\bibnamefont {Watanabe}}, \bibinfo {author}
  {\bibfnamefont {T.}~\bibnamefont {Taniguchi}}, \bibinfo {author}
  {\bibfnamefont {P.}~\bibnamefont {Kim}}, \bibinfo {author} {\bibfnamefont
  {H.}~\bibnamefont {Park}},\ and\ \bibinfo {author} {\bibfnamefont {M.~D.}\
  \bibnamefont {Lukin}},\ }\bibfield  {title} {\bibinfo {title} {Excitons in a
  reconstructed moir{\'e}potential in twisted wse2/wse2 homobilayers},\
  }\href@noop {} {\bibfield  {journal} {\bibinfo  {journal} {Nature Materials}\
  ,\ \bibinfo {pages} {480}} (\bibinfo {year} {2021})}\BibitemShut {NoStop}%
\bibitem [{\citenamefont {Brem}\ \emph {et~al.}(2020)\citenamefont {Brem},
  \citenamefont {Lin}, \citenamefont {Gillen}, \citenamefont {Bauer},
  \citenamefont {Maultzsch}, \citenamefont {Lupton},\ and\ \citenamefont
  {Malic}}]{expt7}%
  \BibitemOpen
  \bibfield  {author} {\bibinfo {author} {\bibfnamefont {S.}~\bibnamefont
  {Brem}}, \bibinfo {author} {\bibfnamefont {K.-Q.}\ \bibnamefont {Lin}},
  \bibinfo {author} {\bibfnamefont {R.}~\bibnamefont {Gillen}}, \bibinfo
  {author} {\bibfnamefont {J.~M.}\ \bibnamefont {Bauer}}, \bibinfo {author}
  {\bibfnamefont {J.}~\bibnamefont {Maultzsch}}, \bibinfo {author}
  {\bibfnamefont {J.~M.}\ \bibnamefont {Lupton}},\ and\ \bibinfo {author}
  {\bibfnamefont {E.}~\bibnamefont {Malic}},\ }\bibfield  {title} {\bibinfo
  {title} {Hybridized intervalley moiré excitons and flat bands in twisted
  wse2 bilayers},\ }\href@noop {} {\bibfield  {journal} {\bibinfo  {journal}
  {Nanoscale}\ }\textbf {\bibinfo {volume} {12}},\ \bibinfo {pages} {11088}
  (\bibinfo {year} {2020})}\BibitemShut {NoStop}%
\bibitem [{\citenamefont {Scuri}\ \emph {et~al.}(2020)\citenamefont {Scuri},
  \citenamefont {Andersen}, \citenamefont {Zhou}, \citenamefont {Wild},
  \citenamefont {Sung}, \citenamefont {Gelly}, \citenamefont {B\'erub\'e},
  \citenamefont {Heo}, \citenamefont {Shao}, \citenamefont {Joe}, \citenamefont
  {Mier~Valdivia}, \citenamefont {Taniguchi}, \citenamefont {Watanabe},
  \citenamefont {Lon\ifmmode~\check{c}\else \v{c}\fi{}ar}, \citenamefont {Kim},
  \citenamefont {Lukin},\ and\ \citenamefont {Park}}]{expt8}%
  \BibitemOpen
  \bibfield  {author} {\bibinfo {author} {\bibfnamefont {G.}~\bibnamefont
  {Scuri}}, \bibinfo {author} {\bibfnamefont {T.~I.}\ \bibnamefont {Andersen}},
  \bibinfo {author} {\bibfnamefont {Y.}~\bibnamefont {Zhou}}, \bibinfo {author}
  {\bibfnamefont {D.~S.}\ \bibnamefont {Wild}}, \bibinfo {author}
  {\bibfnamefont {J.}~\bibnamefont {Sung}}, \bibinfo {author} {\bibfnamefont
  {R.~J.}\ \bibnamefont {Gelly}}, \bibinfo {author} {\bibfnamefont
  {D.}~\bibnamefont {B\'erub\'e}}, \bibinfo {author} {\bibfnamefont
  {H.}~\bibnamefont {Heo}}, \bibinfo {author} {\bibfnamefont {L.}~\bibnamefont
  {Shao}}, \bibinfo {author} {\bibfnamefont {A.~Y.}\ \bibnamefont {Joe}},
  \bibinfo {author} {\bibfnamefont {A.~M.}\ \bibnamefont {Mier~Valdivia}},
  \bibinfo {author} {\bibfnamefont {T.}~\bibnamefont {Taniguchi}}, \bibinfo
  {author} {\bibfnamefont {K.}~\bibnamefont {Watanabe}}, \bibinfo {author}
  {\bibfnamefont {M.}~\bibnamefont {Lon\ifmmode~\check{c}\else \v{c}\fi{}ar}},
  \bibinfo {author} {\bibfnamefont {P.}~\bibnamefont {Kim}}, \bibinfo {author}
  {\bibfnamefont {M.~D.}\ \bibnamefont {Lukin}},\ and\ \bibinfo {author}
  {\bibfnamefont {H.}~\bibnamefont {Park}},\ }\bibfield  {title} {\bibinfo
  {title} {Electrically tunable valley dynamics in twisted
  ${\mathrm{wse}}_{2}/{\mathrm{wse}}_{2}$ bilayers},\ }\href@noop {} {\bibfield
   {journal} {\bibinfo  {journal} {Phys. Rev. Lett.}\ }\textbf {\bibinfo
  {volume} {124}},\ \bibinfo {pages} {217403} (\bibinfo {year}
  {2020})}\BibitemShut {NoStop}%
\bibitem [{\citenamefont {Mahdikhanysarvejahany}\ \emph
  {et~al.}(2021)\citenamefont {Mahdikhanysarvejahany}, \citenamefont {Shanks},
  \citenamefont {Muccianti}, \citenamefont {Badada}, \citenamefont {Idi},
  \citenamefont {Alfrey}, \citenamefont {Raglow}, \citenamefont {Koehler},
  \citenamefont {Mandrus}, \citenamefont {Taniguchi}, \citenamefont {Watanabe},
  \citenamefont {Monti}, \citenamefont {Yu}, \citenamefont {LeRoy},\ and\
  \citenamefont {Schaibley}}]{leroy2}%
  \BibitemOpen
  \bibfield  {author} {\bibinfo {author} {\bibfnamefont {F.}~\bibnamefont
  {Mahdikhanysarvejahany}}, \bibinfo {author} {\bibfnamefont {D.~N.}\
  \bibnamefont {Shanks}}, \bibinfo {author} {\bibfnamefont {C.}~\bibnamefont
  {Muccianti}}, \bibinfo {author} {\bibfnamefont {B.~H.}\ \bibnamefont
  {Badada}}, \bibinfo {author} {\bibfnamefont {I.}~\bibnamefont {Idi}},
  \bibinfo {author} {\bibfnamefont {A.}~\bibnamefont {Alfrey}}, \bibinfo
  {author} {\bibfnamefont {S.}~\bibnamefont {Raglow}}, \bibinfo {author}
  {\bibfnamefont {M.~R.}\ \bibnamefont {Koehler}}, \bibinfo {author}
  {\bibfnamefont {D.~G.}\ \bibnamefont {Mandrus}}, \bibinfo {author}
  {\bibfnamefont {T.}~\bibnamefont {Taniguchi}}, \bibinfo {author}
  {\bibfnamefont {K.}~\bibnamefont {Watanabe}}, \bibinfo {author}
  {\bibfnamefont {O.~L.~A.}\ \bibnamefont {Monti}}, \bibinfo {author}
  {\bibfnamefont {H.}~\bibnamefont {Yu}}, \bibinfo {author} {\bibfnamefont
  {B.~J.}\ \bibnamefont {LeRoy}},\ and\ \bibinfo {author} {\bibfnamefont
  {J.~R.}\ \bibnamefont {Schaibley}},\ }\bibfield  {title} {\bibinfo {title}
  {Temperature dependent moir{\'e}trapping of interlayer excitons in mose2-wse2
  heterostructures},\ }\href@noop {} {\bibfield  {journal} {\bibinfo  {journal}
  {npj 2D Materials and Applications}\ }\textbf {\bibinfo {volume} {5}},\
  \bibinfo {pages} {67} (\bibinfo {year} {2021})}\BibitemShut {NoStop}%
\bibitem [{\citenamefont {Naik}\ \emph {et~al.}(2022)\citenamefont {Naik},
  \citenamefont {Regan}, \citenamefont {Zhang}, \citenamefont {Chan},
  \citenamefont {Li}, \citenamefont {Wang}, \citenamefont {Yoon}, \citenamefont
  {Ong}, \citenamefont {Zhao}, \citenamefont {Zhao} \emph
  {et~al.}}]{naik2022intralayer}%
  \BibitemOpen
  \bibfield  {author} {\bibinfo {author} {\bibfnamefont {M.~H.}\ \bibnamefont
  {Naik}}, \bibinfo {author} {\bibfnamefont {E.~C.}\ \bibnamefont {Regan}},
  \bibinfo {author} {\bibfnamefont {Z.}~\bibnamefont {Zhang}}, \bibinfo
  {author} {\bibfnamefont {Y.-H.}\ \bibnamefont {Chan}}, \bibinfo {author}
  {\bibfnamefont {Z.}~\bibnamefont {Li}}, \bibinfo {author} {\bibfnamefont
  {D.}~\bibnamefont {Wang}}, \bibinfo {author} {\bibfnamefont {Y.}~\bibnamefont
  {Yoon}}, \bibinfo {author} {\bibfnamefont {C.~S.}\ \bibnamefont {Ong}},
  \bibinfo {author} {\bibfnamefont {W.}~\bibnamefont {Zhao}}, \bibinfo {author}
  {\bibfnamefont {S.}~\bibnamefont {Zhao}}, \emph {et~al.},\ }\bibfield
  {title} {\bibinfo {title} {Intralayer charge-transfer moir{\'e} excitons in
  van der waals superlattices},\ }\href@noop {} {\bibfield  {journal} {\bibinfo
   {journal} {Nature}\ }\textbf {\bibinfo {volume} {609}},\ \bibinfo {pages}
  {52} (\bibinfo {year} {2022})}\BibitemShut {NoStop}%
\bibitem [{\citenamefont {Xiong}\ \emph {et~al.}(2023)\citenamefont {Xiong},
  \citenamefont {Nie}, \citenamefont {Brantly}, \citenamefont {Hays},
  \citenamefont {Sailus}, \citenamefont {Watanabe}, \citenamefont {Taniguchi},
  \citenamefont {Tongay},\ and\ \citenamefont {Jin}}]{xiong2023correlated}%
  \BibitemOpen
  \bibfield  {author} {\bibinfo {author} {\bibfnamefont {R.}~\bibnamefont
  {Xiong}}, \bibinfo {author} {\bibfnamefont {J.~H.}\ \bibnamefont {Nie}},
  \bibinfo {author} {\bibfnamefont {S.~L.}\ \bibnamefont {Brantly}}, \bibinfo
  {author} {\bibfnamefont {P.}~\bibnamefont {Hays}}, \bibinfo {author}
  {\bibfnamefont {R.}~\bibnamefont {Sailus}}, \bibinfo {author} {\bibfnamefont
  {K.}~\bibnamefont {Watanabe}}, \bibinfo {author} {\bibfnamefont
  {T.}~\bibnamefont {Taniguchi}}, \bibinfo {author} {\bibfnamefont
  {S.}~\bibnamefont {Tongay}},\ and\ \bibinfo {author} {\bibfnamefont
  {C.}~\bibnamefont {Jin}},\ }\bibfield  {title} {\bibinfo {title} {Correlated
  insulator of excitons in wse2/ws2 moir{\'e} superlattices},\ }\href@noop {}
  {\bibfield  {journal} {\bibinfo  {journal} {Science}\ }\textbf {\bibinfo
  {volume} {380}},\ \bibinfo {pages} {860} (\bibinfo {year}
  {2023})}\BibitemShut {NoStop}%
\bibitem [{\citenamefont {Brotons-Gisbert}\ \emph {et~al.}(2024)\citenamefont
  {Brotons-Gisbert}, \citenamefont {Gerardot}, \citenamefont {Holleitner},\
  and\ \citenamefont {Wurstbauer}}]{brotons2024interlayer}%
  \BibitemOpen
  \bibfield  {author} {\bibinfo {author} {\bibfnamefont {M.}~\bibnamefont
  {Brotons-Gisbert}}, \bibinfo {author} {\bibfnamefont {B.~D.}\ \bibnamefont
  {Gerardot}}, \bibinfo {author} {\bibfnamefont {A.~W.}\ \bibnamefont
  {Holleitner}},\ and\ \bibinfo {author} {\bibfnamefont {U.}~\bibnamefont
  {Wurstbauer}},\ }\bibfield  {title} {\bibinfo {title} {Interlayer and
  moir{\'e} excitons in atomically thin double layers: From individual quantum
  emitters to degenerate ensembles},\ }\href@noop {} {\bibfield  {journal}
  {\bibinfo  {journal} {MRS bulletin}\ }\textbf {\bibinfo {volume} {49}},\
  \bibinfo {pages} {914} (\bibinfo {year} {2024})}\BibitemShut {NoStop}%
\bibitem [{\citenamefont {Naik}\ and\ \citenamefont {Jain}(2018)}]{naik1}%
  \BibitemOpen
  \bibfield  {author} {\bibinfo {author} {\bibfnamefont {M.~H.}\ \bibnamefont
  {Naik}}\ and\ \bibinfo {author} {\bibfnamefont {M.}~\bibnamefont {Jain}},\
  }\bibfield  {title} {\bibinfo {title} {Ultraflatbands and shear solitons in
  moir\'e patterns of twisted bilayer transition metal dichalcogenides},\
  }\href@noop {} {\bibfield  {journal} {\bibinfo  {journal} {Phys. Rev. Lett.}\
  }\textbf {\bibinfo {volume} {121}},\ \bibinfo {pages} {266401} (\bibinfo
  {year} {2018})}\BibitemShut {NoStop}%
\bibitem [{\citenamefont {Naik}\ \emph {et~al.}(2020)\citenamefont {Naik},
  \citenamefont {Kundu}, \citenamefont {Maity},\ and\ \citenamefont
  {Jain}}]{naik2}%
  \BibitemOpen
  \bibfield  {author} {\bibinfo {author} {\bibfnamefont {M.~H.}\ \bibnamefont
  {Naik}}, \bibinfo {author} {\bibfnamefont {S.}~\bibnamefont {Kundu}},
  \bibinfo {author} {\bibfnamefont {I.}~\bibnamefont {Maity}},\ and\ \bibinfo
  {author} {\bibfnamefont {M.}~\bibnamefont {Jain}},\ }\bibfield  {title}
  {\bibinfo {title} {Origin and evolution of ultraflat bands in twisted bilayer
  transition metal dichalcogenides: Realization of triangular quantum dots},\
  }\href@noop {} {\bibfield  {journal} {\bibinfo  {journal} {Phys. Rev. B}\
  }\textbf {\bibinfo {volume} {102}},\ \bibinfo {pages} {075413} (\bibinfo
  {year} {2020})}\BibitemShut {NoStop}%
\bibitem [{\citenamefont {Wu}\ \emph {et~al.}(2019)\citenamefont {Wu},
  \citenamefont {Lovorn}, \citenamefont {Tutuc}, \citenamefont {Martin},\ and\
  \citenamefont {MacDonald}}]{continm}%
  \BibitemOpen
  \bibfield  {author} {\bibinfo {author} {\bibfnamefont {F.}~\bibnamefont
  {Wu}}, \bibinfo {author} {\bibfnamefont {T.}~\bibnamefont {Lovorn}}, \bibinfo
  {author} {\bibfnamefont {E.}~\bibnamefont {Tutuc}}, \bibinfo {author}
  {\bibfnamefont {I.}~\bibnamefont {Martin}},\ and\ \bibinfo {author}
  {\bibfnamefont {A.~H.}\ \bibnamefont {MacDonald}},\ }\bibfield  {title}
  {\bibinfo {title} {Topological insulators in twisted transition metal
  dichalcogenide homobilayers},\ }\href@noop {} {\bibfield  {journal} {\bibinfo
   {journal} {Phys. Rev. Lett.}\ }\textbf {\bibinfo {volume} {122}},\ \bibinfo
  {pages} {086402} (\bibinfo {year} {2019})}\BibitemShut {NoStop}%
\bibitem [{\citenamefont {Pan}\ \emph {et~al.}(2020)\citenamefont {Pan},
  \citenamefont {Wu},\ and\ \citenamefont {Das~Sarma}}]{sds1}%
  \BibitemOpen
  \bibfield  {author} {\bibinfo {author} {\bibfnamefont {H.}~\bibnamefont
  {Pan}}, \bibinfo {author} {\bibfnamefont {F.}~\bibnamefont {Wu}},\ and\
  \bibinfo {author} {\bibfnamefont {S.}~\bibnamefont {Das~Sarma}},\ }\bibfield
  {title} {\bibinfo {title} {Band topology, hubbard model, heisenberg model,
  and dzyaloshinskii-moriya interaction in twisted bilayer
  ${\mathrm{wse}}_{2}$},\ }\href@noop {} {\bibfield  {journal} {\bibinfo
  {journal} {Phys. Rev. Research}\ }\textbf {\bibinfo {volume} {2}},\ \bibinfo
  {pages} {033087} (\bibinfo {year} {2020})}\BibitemShut {NoStop}%
\bibitem [{\citenamefont {Zhang}\ \emph {et~al.}(2021)\citenamefont {Zhang},
  \citenamefont {Liu},\ and\ \citenamefont {Fu}}]{fu1}%
  \BibitemOpen
  \bibfield  {author} {\bibinfo {author} {\bibfnamefont {Y.}~\bibnamefont
  {Zhang}}, \bibinfo {author} {\bibfnamefont {T.}~\bibnamefont {Liu}},\ and\
  \bibinfo {author} {\bibfnamefont {L.}~\bibnamefont {Fu}},\ }\bibfield
  {title} {\bibinfo {title} {Electronic structures, charge transfer, and charge
  order in twisted transition metal dichalcogenide bilayers},\ }\href@noop {}
  {\bibfield  {journal} {\bibinfo  {journal} {Phys. Rev. B}\ }\textbf {\bibinfo
  {volume} {103}},\ \bibinfo {pages} {155142} (\bibinfo {year}
  {2021})}\BibitemShut {NoStop}%
\bibitem [{\citenamefont {Maity}\ \emph {et~al.}(2021)\citenamefont {Maity},
  \citenamefont {Maiti}, \citenamefont {Krishnamurthy},\ and\ \citenamefont
  {Jain}}]{maity1}%
  \BibitemOpen
  \bibfield  {author} {\bibinfo {author} {\bibfnamefont {I.}~\bibnamefont
  {Maity}}, \bibinfo {author} {\bibfnamefont {P.~K.}\ \bibnamefont {Maiti}},
  \bibinfo {author} {\bibfnamefont {H.~R.}\ \bibnamefont {Krishnamurthy}},\
  and\ \bibinfo {author} {\bibfnamefont {M.}~\bibnamefont {Jain}},\ }\bibfield
  {title} {\bibinfo {title} {Reconstruction of moir\'e lattices in twisted
  transition metal dichalcogenide bilayers},\ }\href@noop {} {\bibfield
  {journal} {\bibinfo  {journal} {Phys. Rev. B}\ }\textbf {\bibinfo {volume}
  {103}},\ \bibinfo {pages} {L121102} (\bibinfo {year} {2021})}\BibitemShut
  {NoStop}%
\bibitem [{\citenamefont {Maity}\ \emph {et~al.}(2020)\citenamefont {Maity},
  \citenamefont {Naik}, \citenamefont {Maiti}, \citenamefont {Krishnamurthy},\
  and\ \citenamefont {Jain}}]{maity2}%
  \BibitemOpen
  \bibfield  {author} {\bibinfo {author} {\bibfnamefont {I.}~\bibnamefont
  {Maity}}, \bibinfo {author} {\bibfnamefont {M.~H.}\ \bibnamefont {Naik}},
  \bibinfo {author} {\bibfnamefont {P.~K.}\ \bibnamefont {Maiti}}, \bibinfo
  {author} {\bibfnamefont {H.~R.}\ \bibnamefont {Krishnamurthy}},\ and\
  \bibinfo {author} {\bibfnamefont {M.}~\bibnamefont {Jain}},\ }\bibfield
  {title} {\bibinfo {title} {Phonons in twisted transition-metal dichalcogenide
  bilayers: Ultrasoft phasons and a transition from a superlubric to a pinned
  phase},\ }\href@noop {} {\bibfield  {journal} {\bibinfo  {journal} {Phys.
  Rev. Research}\ }\textbf {\bibinfo {volume} {2}},\ \bibinfo {pages} {013335}
  (\bibinfo {year} {2020})}\BibitemShut {NoStop}%
\bibitem [{\citenamefont {Xian}\ \emph {et~al.}(2021)\citenamefont {Xian},
  \citenamefont {Claassen}, \citenamefont {Kiese}, \citenamefont {Scherer},
  \citenamefont {Trebst}, \citenamefont {Kennes},\ and\ \citenamefont
  {Rubio}}]{rubio}%
  \BibitemOpen
  \bibfield  {author} {\bibinfo {author} {\bibfnamefont {L.}~\bibnamefont
  {Xian}}, \bibinfo {author} {\bibfnamefont {M.}~\bibnamefont {Claassen}},
  \bibinfo {author} {\bibfnamefont {D.}~\bibnamefont {Kiese}}, \bibinfo
  {author} {\bibfnamefont {M.~M.}\ \bibnamefont {Scherer}}, \bibinfo {author}
  {\bibfnamefont {S.}~\bibnamefont {Trebst}}, \bibinfo {author} {\bibfnamefont
  {D.~M.}\ \bibnamefont {Kennes}},\ and\ \bibinfo {author} {\bibfnamefont
  {A.}~\bibnamefont {Rubio}},\ }\bibfield  {title} {\bibinfo {title}
  {Realization of nearly dispersionless bands with strong orbital anisotropy
  from destructive interference in twisted bilayer mos2},\ }\href@noop {}
  {\bibfield  {journal} {\bibinfo  {journal} {Nature communications}\ }\textbf
  {\bibinfo {volume} {12}},\ \bibinfo {pages} {5644} (\bibinfo {year}
  {2021})}\BibitemShut {NoStop}%
\bibitem [{\citenamefont {Angeli}\ and\ \citenamefont
  {MacDonald}(2021)}]{gammavalley}%
  \BibitemOpen
  \bibfield  {author} {\bibinfo {author} {\bibfnamefont {M.}~\bibnamefont
  {Angeli}}\ and\ \bibinfo {author} {\bibfnamefont {A.~H.}\ \bibnamefont
  {MacDonald}},\ }\bibfield  {title} {\bibinfo {title} {$\gamma$ valley
  transition metal dichalcogenide moir{\'e} bands},\ }\href@noop {} {\bibfield
  {journal} {\bibinfo  {journal} {Proceedings of the National Academy of
  Sciences}\ }\textbf {\bibinfo {volume} {118}},\ \bibinfo {pages}
  {e2021826118} (\bibinfo {year} {2021})}\BibitemShut {NoStop}%
\bibitem [{\citenamefont {Kundu}\ \emph {et~al.}(2022)\citenamefont {Kundu},
  \citenamefont {Naik}, \citenamefont {Krishnamurthy},\ and\ \citenamefont
  {Jain}}]{kundu1}%
  \BibitemOpen
  \bibfield  {author} {\bibinfo {author} {\bibfnamefont {S.}~\bibnamefont
  {Kundu}}, \bibinfo {author} {\bibfnamefont {M.~H.}\ \bibnamefont {Naik}},
  \bibinfo {author} {\bibfnamefont {H.}~\bibnamefont {Krishnamurthy}},\ and\
  \bibinfo {author} {\bibfnamefont {M.}~\bibnamefont {Jain}},\ }\bibfield
  {title} {\bibinfo {title} {Moir{\'e} induced topology and flat bands in
  twisted bilayer wse 2: A first-principles study},\ }\href@noop {} {\bibfield
  {journal} {\bibinfo  {journal} {Physical Review B}\ }\textbf {\bibinfo
  {volume} {105}},\ \bibinfo {pages} {L081108} (\bibinfo {year}
  {2022})}\BibitemShut {NoStop}%
\bibitem [{\citenamefont {Zhang}\ \emph {et~al.}(2020)\citenamefont {Zhang},
  \citenamefont {Wang}, \citenamefont {Watanabe}, \citenamefont {Taniguchi},
  \citenamefont {Ueno}, \citenamefont {Tutuc},\ and\ \citenamefont
  {LeRoy}}]{wse2_expt1}%
  \BibitemOpen
  \bibfield  {author} {\bibinfo {author} {\bibfnamefont {Z.}~\bibnamefont
  {Zhang}}, \bibinfo {author} {\bibfnamefont {Y.}~\bibnamefont {Wang}},
  \bibinfo {author} {\bibfnamefont {K.}~\bibnamefont {Watanabe}}, \bibinfo
  {author} {\bibfnamefont {T.}~\bibnamefont {Taniguchi}}, \bibinfo {author}
  {\bibfnamefont {K.}~\bibnamefont {Ueno}}, \bibinfo {author} {\bibfnamefont
  {E.}~\bibnamefont {Tutuc}},\ and\ \bibinfo {author} {\bibfnamefont {B.~J.}\
  \bibnamefont {LeRoy}},\ }\bibfield  {title} {\bibinfo {title} {Flat bands in
  twisted bilayer transition metal dichalcogenides},\ }\href@noop {} {\bibfield
   {journal} {\bibinfo  {journal} {Nature Physics}\ }\textbf {\bibinfo {volume}
  {16}},\ \bibinfo {pages} {1093} (\bibinfo {year} {2020})}\BibitemShut
  {NoStop}%
\bibitem [{\citenamefont {Soriano}\ and\ \citenamefont {Lado}(2020)}]{lado1}%
  \BibitemOpen
  \bibfield  {author} {\bibinfo {author} {\bibfnamefont {D.}~\bibnamefont
  {Soriano}}\ and\ \bibinfo {author} {\bibfnamefont {J.~L.}\ \bibnamefont
  {Lado}},\ }\bibfield  {title} {\bibinfo {title} {Exchange-bias controlled
  correlations in magnetically encapsulated twisted van der waals
  dichalcogenides},\ }\href@noop {} {\bibfield  {journal} {\bibinfo  {journal}
  {Journal of Physics D: Applied Physics}\ }\textbf {\bibinfo {volume} {53}},\
  \bibinfo {pages} {474001} (\bibinfo {year} {2020})}\BibitemShut {NoStop}%
\bibitem [{\citenamefont {Weston}\ \emph {et~al.}(2020)\citenamefont {Weston},
  \citenamefont {Zou}, \citenamefont {Enaldiev}, \citenamefont {Summerfield},
  \citenamefont {Clark}, \citenamefont {Z{\'o}lyomi}, \citenamefont {Graham},
  \citenamefont {Yelgel}, \citenamefont {Magorrian}, \citenamefont {Zhou},
  \citenamefont {Zultak}, \citenamefont {Hopkinson}, \citenamefont {Barinov},
  \citenamefont {Bointon}, \citenamefont {Kretinin}, \citenamefont {Wilson},
  \citenamefont {Beton}, \citenamefont {Fal'ko}, \citenamefont {Haigh},\ and\
  \citenamefont {Gorbachev}}]{expt9}%
  \BibitemOpen
  \bibfield  {author} {\bibinfo {author} {\bibfnamefont {A.}~\bibnamefont
  {Weston}}, \bibinfo {author} {\bibfnamefont {Y.}~\bibnamefont {Zou}},
  \bibinfo {author} {\bibfnamefont {V.}~\bibnamefont {Enaldiev}}, \bibinfo
  {author} {\bibfnamefont {A.}~\bibnamefont {Summerfield}}, \bibinfo {author}
  {\bibfnamefont {N.}~\bibnamefont {Clark}}, \bibinfo {author} {\bibfnamefont
  {V.}~\bibnamefont {Z{\'o}lyomi}}, \bibinfo {author} {\bibfnamefont
  {A.}~\bibnamefont {Graham}}, \bibinfo {author} {\bibfnamefont
  {C.}~\bibnamefont {Yelgel}}, \bibinfo {author} {\bibfnamefont
  {S.}~\bibnamefont {Magorrian}}, \bibinfo {author} {\bibfnamefont
  {M.}~\bibnamefont {Zhou}}, \bibinfo {author} {\bibfnamefont {J.}~\bibnamefont
  {Zultak}}, \bibinfo {author} {\bibfnamefont {D.}~\bibnamefont {Hopkinson}},
  \bibinfo {author} {\bibfnamefont {A.}~\bibnamefont {Barinov}}, \bibinfo
  {author} {\bibfnamefont {T.~H.}\ \bibnamefont {Bointon}}, \bibinfo {author}
  {\bibfnamefont {A.}~\bibnamefont {Kretinin}}, \bibinfo {author}
  {\bibfnamefont {N.~R.}\ \bibnamefont {Wilson}}, \bibinfo {author}
  {\bibfnamefont {P.~H.}\ \bibnamefont {Beton}}, \bibinfo {author}
  {\bibfnamefont {V.~I.}\ \bibnamefont {Fal'ko}}, \bibinfo {author}
  {\bibfnamefont {S.~J.}\ \bibnamefont {Haigh}},\ and\ \bibinfo {author}
  {\bibfnamefont {R.}~\bibnamefont {Gorbachev}},\ }\bibfield  {title} {\bibinfo
  {title} {Atomic reconstruction in twisted bilayers of transition metal
  dichalcogenides},\ }\href {https://doi.org/10.1038/s41565-020-0682-9}
  {\bibfield  {journal} {\bibinfo  {journal} {Nature Nanotechnology}\ }\textbf
  {\bibinfo {volume} {15}},\ \bibinfo {pages} {592} (\bibinfo {year}
  {2020})}\BibitemShut {NoStop}%
\bibitem [{\citenamefont {Halbertal}\ \emph {et~al.}(2021)\citenamefont
  {Halbertal}, \citenamefont {Finney}, \citenamefont {Sunku}, \citenamefont
  {Kerelsky}, \citenamefont {Rubio-Verd{\'u}}, \citenamefont {Shabani},
  \citenamefont {Xian}, \citenamefont {Carr}, \citenamefont {Chen},
  \citenamefont {Zhang}, \citenamefont {Wang}, \citenamefont
  {Gonzalez-Acevedo}, \citenamefont {McLeod}, \citenamefont {Rhodes},
  \citenamefont {Watanabe}, \citenamefont {Taniguchi}, \citenamefont {Kaxiras},
  \citenamefont {Dean}, \citenamefont {Hone}, \citenamefont {Pasupathy},
  \citenamefont {Kennes}, \citenamefont {Rubio},\ and\ \citenamefont
  {Basov}}]{expt13}%
  \BibitemOpen
  \bibfield  {author} {\bibinfo {author} {\bibfnamefont {D.}~\bibnamefont
  {Halbertal}}, \bibinfo {author} {\bibfnamefont {N.~R.}\ \bibnamefont
  {Finney}}, \bibinfo {author} {\bibfnamefont {S.~S.}\ \bibnamefont {Sunku}},
  \bibinfo {author} {\bibfnamefont {A.}~\bibnamefont {Kerelsky}}, \bibinfo
  {author} {\bibfnamefont {C.}~\bibnamefont {Rubio-Verd{\'u}}}, \bibinfo
  {author} {\bibfnamefont {S.}~\bibnamefont {Shabani}}, \bibinfo {author}
  {\bibfnamefont {L.}~\bibnamefont {Xian}}, \bibinfo {author} {\bibfnamefont
  {S.}~\bibnamefont {Carr}}, \bibinfo {author} {\bibfnamefont {S.}~\bibnamefont
  {Chen}}, \bibinfo {author} {\bibfnamefont {C.}~\bibnamefont {Zhang}},
  \bibinfo {author} {\bibfnamefont {L.}~\bibnamefont {Wang}}, \bibinfo {author}
  {\bibfnamefont {D.}~\bibnamefont {Gonzalez-Acevedo}}, \bibinfo {author}
  {\bibfnamefont {A.~S.}\ \bibnamefont {McLeod}}, \bibinfo {author}
  {\bibfnamefont {D.}~\bibnamefont {Rhodes}}, \bibinfo {author} {\bibfnamefont
  {K.}~\bibnamefont {Watanabe}}, \bibinfo {author} {\bibfnamefont
  {T.}~\bibnamefont {Taniguchi}}, \bibinfo {author} {\bibfnamefont
  {E.}~\bibnamefont {Kaxiras}}, \bibinfo {author} {\bibfnamefont {C.~R.}\
  \bibnamefont {Dean}}, \bibinfo {author} {\bibfnamefont {J.~C.}\ \bibnamefont
  {Hone}}, \bibinfo {author} {\bibfnamefont {A.~N.}\ \bibnamefont {Pasupathy}},
  \bibinfo {author} {\bibfnamefont {D.~M.}\ \bibnamefont {Kennes}}, \bibinfo
  {author} {\bibfnamefont {A.}~\bibnamefont {Rubio}},\ and\ \bibinfo {author}
  {\bibfnamefont {D.~N.}\ \bibnamefont {Basov}},\ }\bibfield  {title} {\bibinfo
  {title} {Moir{\'e}metrology of energy landscapes in van der waals
  heterostructures},\ }\href@noop {} {\bibfield  {journal} {\bibinfo  {journal}
  {Nature Communications}\ }\textbf {\bibinfo {volume} {12}},\ \bibinfo {pages}
  {242} (\bibinfo {year} {2021})}\BibitemShut {NoStop}%
\bibitem [{\citenamefont {Kennes}\ \emph {et~al.}(2021)\citenamefont {Kennes},
  \citenamefont {Claassen}, \citenamefont {Xian}, \citenamefont {Georges},
  \citenamefont {Millis}, \citenamefont {Hone}, \citenamefont {Dean},
  \citenamefont {Basov}, \citenamefont {Pasupathy},\ and\ \citenamefont
  {Rubio}}]{rubio2}%
  \BibitemOpen
  \bibfield  {author} {\bibinfo {author} {\bibfnamefont {D.~M.}\ \bibnamefont
  {Kennes}}, \bibinfo {author} {\bibfnamefont {M.}~\bibnamefont {Claassen}},
  \bibinfo {author} {\bibfnamefont {L.}~\bibnamefont {Xian}}, \bibinfo {author}
  {\bibfnamefont {A.}~\bibnamefont {Georges}}, \bibinfo {author} {\bibfnamefont
  {A.~J.}\ \bibnamefont {Millis}}, \bibinfo {author} {\bibfnamefont
  {J.}~\bibnamefont {Hone}}, \bibinfo {author} {\bibfnamefont {C.~R.}\
  \bibnamefont {Dean}}, \bibinfo {author} {\bibfnamefont {D.~N.}\ \bibnamefont
  {Basov}}, \bibinfo {author} {\bibfnamefont {A.~N.}\ \bibnamefont
  {Pasupathy}},\ and\ \bibinfo {author} {\bibfnamefont {A.}~\bibnamefont
  {Rubio}},\ }\bibfield  {title} {\bibinfo {title} {Moir{\'e}heterostructures
  as a condensed-matter quantum simulator},\ }\href@noop {} {\bibfield
  {journal} {\bibinfo  {journal} {Nature Physics}\ }\textbf {\bibinfo {volume}
  {17}},\ \bibinfo {pages} {155} (\bibinfo {year} {2021})}\BibitemShut
  {NoStop}%
\bibitem [{\citenamefont {Wu}\ \emph {et~al.}(2018)\citenamefont {Wu},
  \citenamefont {Lovorn}, \citenamefont {Tutuc},\ and\ \citenamefont
  {MacDonald}}]{macdonald1}%
  \BibitemOpen
  \bibfield  {author} {\bibinfo {author} {\bibfnamefont {F.}~\bibnamefont
  {Wu}}, \bibinfo {author} {\bibfnamefont {T.}~\bibnamefont {Lovorn}}, \bibinfo
  {author} {\bibfnamefont {E.}~\bibnamefont {Tutuc}},\ and\ \bibinfo {author}
  {\bibfnamefont {A.~H.}\ \bibnamefont {MacDonald}},\ }\bibfield  {title}
  {\bibinfo {title} {Hubbard model physics in transition metal dichalcogenide
  moir\'e bands},\ }\href@noop {} {\bibfield  {journal} {\bibinfo  {journal}
  {Phys. Rev. Lett.}\ }\textbf {\bibinfo {volume} {121}},\ \bibinfo {pages}
  {026402} (\bibinfo {year} {2018})}\BibitemShut {NoStop}%
\bibitem [{\citenamefont {Vitale}\ \emph {et~al.}(2021)\citenamefont {Vitale},
  \citenamefont {Atalar}, \citenamefont {Mostofi},\ and\ \citenamefont
  {Lischner}}]{vitale}%
  \BibitemOpen
  \bibfield  {author} {\bibinfo {author} {\bibfnamefont {V.}~\bibnamefont
  {Vitale}}, \bibinfo {author} {\bibfnamefont {K.}~\bibnamefont {Atalar}},
  \bibinfo {author} {\bibfnamefont {A.~A.}\ \bibnamefont {Mostofi}},\ and\
  \bibinfo {author} {\bibfnamefont {J.}~\bibnamefont {Lischner}},\ }\bibfield
  {title} {\bibinfo {title} {Flat band properties of twisted transition metal
  dichalcogenide homo-and heterobilayers of mos2, mose2, ws2 and wse2},\
  }\href@noop {} {\bibfield  {journal} {\bibinfo  {journal} {2D Materials}\
  }\textbf {\bibinfo {volume} {8}},\ \bibinfo {pages} {045010} (\bibinfo {year}
  {2021})}\BibitemShut {NoStop}%
\bibitem [{\citenamefont {Sinner}\ \emph {et~al.}(2023)\citenamefont {Sinner},
  \citenamefont {Pantale{\'o}n},\ and\ \citenamefont
  {Guinea}}]{sinner2023strain}%
  \BibitemOpen
  \bibfield  {author} {\bibinfo {author} {\bibfnamefont {A.}~\bibnamefont
  {Sinner}}, \bibinfo {author} {\bibfnamefont {P.~A.}\ \bibnamefont
  {Pantale{\'o}n}},\ and\ \bibinfo {author} {\bibfnamefont {F.}~\bibnamefont
  {Guinea}},\ }\bibfield  {title} {\bibinfo {title} {Strain-induced quasi-1d
  channels in twisted moir{\'e} lattices},\ }\href@noop {} {\bibfield
  {journal} {\bibinfo  {journal} {Physical Review Letters}\ }\textbf {\bibinfo
  {volume} {131}},\ \bibinfo {pages} {166402} (\bibinfo {year}
  {2023})}\BibitemShut {NoStop}%
\bibitem [{\citenamefont {Engelke}\ \emph {et~al.}(2023)\citenamefont
  {Engelke}, \citenamefont {Yoo}, \citenamefont {Carr}, \citenamefont {Xu},
  \citenamefont {Cazeaux}, \citenamefont {Allen}, \citenamefont {Valdivia},
  \citenamefont {Luskin}, \citenamefont {Kaxiras}, \citenamefont {Kim} \emph
  {et~al.}}]{engelke2023topological}%
  \BibitemOpen
  \bibfield  {author} {\bibinfo {author} {\bibfnamefont {R.}~\bibnamefont
  {Engelke}}, \bibinfo {author} {\bibfnamefont {H.}~\bibnamefont {Yoo}},
  \bibinfo {author} {\bibfnamefont {S.}~\bibnamefont {Carr}}, \bibinfo {author}
  {\bibfnamefont {K.}~\bibnamefont {Xu}}, \bibinfo {author} {\bibfnamefont
  {P.}~\bibnamefont {Cazeaux}}, \bibinfo {author} {\bibfnamefont
  {R.}~\bibnamefont {Allen}}, \bibinfo {author} {\bibfnamefont {A.~M.}\
  \bibnamefont {Valdivia}}, \bibinfo {author} {\bibfnamefont {M.}~\bibnamefont
  {Luskin}}, \bibinfo {author} {\bibfnamefont {E.}~\bibnamefont {Kaxiras}},
  \bibinfo {author} {\bibfnamefont {M.}~\bibnamefont {Kim}}, \emph {et~al.},\
  }\bibfield  {title} {\bibinfo {title} {Topological nature of dislocation
  networks in two-dimensional moir{\'e} materials},\ }\href@noop {} {\bibfield
  {journal} {\bibinfo  {journal} {Physical Review B}\ }\textbf {\bibinfo
  {volume} {107}},\ \bibinfo {pages} {125413} (\bibinfo {year}
  {2023})}\BibitemShut {NoStop}%
\bibitem [{\citenamefont {Escudero}\ \emph {et~al.}(2024)\citenamefont
  {Escudero}, \citenamefont {Sinner}, \citenamefont {Zhan}, \citenamefont
  {Pantale{\'o}n},\ and\ \citenamefont {Guinea}}]{escudero2024designing}%
  \BibitemOpen
  \bibfield  {author} {\bibinfo {author} {\bibfnamefont {F.}~\bibnamefont
  {Escudero}}, \bibinfo {author} {\bibfnamefont {A.}~\bibnamefont {Sinner}},
  \bibinfo {author} {\bibfnamefont {Z.}~\bibnamefont {Zhan}}, \bibinfo {author}
  {\bibfnamefont {P.~A.}\ \bibnamefont {Pantale{\'o}n}},\ and\ \bibinfo
  {author} {\bibfnamefont {F.}~\bibnamefont {Guinea}},\ }\bibfield  {title}
  {\bibinfo {title} {Designing moir{\'e} patterns by strain},\ }\href@noop {}
  {\bibfield  {journal} {\bibinfo  {journal} {Physical Review Research}\
  }\textbf {\bibinfo {volume} {6}},\ \bibinfo {pages} {023203} (\bibinfo {year}
  {2024})}\BibitemShut {NoStop}%
\bibitem [{\citenamefont {K{\"o}gl}\ \emph {et~al.}(2023)\citenamefont
  {K{\"o}gl}, \citenamefont {Soubelet}, \citenamefont {Brotons-Gisbert},
  \citenamefont {Stier}, \citenamefont {Gerardot},\ and\ \citenamefont
  {Finley}}]{kogl2023moire}%
  \BibitemOpen
  \bibfield  {author} {\bibinfo {author} {\bibfnamefont {M.}~\bibnamefont
  {K{\"o}gl}}, \bibinfo {author} {\bibfnamefont {P.}~\bibnamefont {Soubelet}},
  \bibinfo {author} {\bibfnamefont {M.}~\bibnamefont {Brotons-Gisbert}},
  \bibinfo {author} {\bibfnamefont {A.}~\bibnamefont {Stier}}, \bibinfo
  {author} {\bibfnamefont {B.}~\bibnamefont {Gerardot}},\ and\ \bibinfo
  {author} {\bibfnamefont {J.}~\bibnamefont {Finley}},\ }\bibfield  {title}
  {\bibinfo {title} {Moir{\'e} straintronics: a universal platform for
  reconfigurable quantum materials},\ }\href@noop {} {\bibfield  {journal}
  {\bibinfo  {journal} {npj 2D Materials and Applications}\ }\textbf {\bibinfo
  {volume} {7}},\ \bibinfo {pages} {32} (\bibinfo {year} {2023})}\BibitemShut
  {NoStop}%
\bibitem [{\citenamefont {Bi}\ \emph {et~al.}(2019)\citenamefont {Bi},
  \citenamefont {Yuan},\ and\ \citenamefont {Fu}}]{strain1}%
  \BibitemOpen
  \bibfield  {author} {\bibinfo {author} {\bibfnamefont {Z.}~\bibnamefont
  {Bi}}, \bibinfo {author} {\bibfnamefont {N.~F.~Q.}\ \bibnamefont {Yuan}},\
  and\ \bibinfo {author} {\bibfnamefont {L.}~\bibnamefont {Fu}},\ }\bibfield
  {title} {\bibinfo {title} {Designing flat bands by strain},\ }\href@noop {}
  {\bibfield  {journal} {\bibinfo  {journal} {Phys. Rev. B}\ }\textbf {\bibinfo
  {volume} {100}},\ \bibinfo {pages} {035448} (\bibinfo {year}
  {2019})}\BibitemShut {NoStop}%
\bibitem [{\citenamefont {Manna\"{\i}}\ and\ \citenamefont
  {Haddad}(2021)}]{strain2}%
  \BibitemOpen
  \bibfield  {author} {\bibinfo {author} {\bibfnamefont {M.}~\bibnamefont
  {Manna\"{\i}}}\ and\ \bibinfo {author} {\bibfnamefont {S.}~\bibnamefont
  {Haddad}},\ }\bibfield  {title} {\bibinfo {title} {Twistronics versus
  straintronics in twisted bilayers of graphene and transition metal
  dichalcogenides},\ }\href@noop {} {\bibfield  {journal} {\bibinfo  {journal}
  {Phys. Rev. B}\ }\textbf {\bibinfo {volume} {103}},\ \bibinfo {pages}
  {L201112} (\bibinfo {year} {2021})}\BibitemShut {NoStop}%
\bibitem [{\citenamefont {Hubbard}\ and\ \citenamefont
  {Torrance}(1981)}]{ionichubbard0}%
  \BibitemOpen
  \bibfield  {author} {\bibinfo {author} {\bibfnamefont {J.}~\bibnamefont
  {Hubbard}}\ and\ \bibinfo {author} {\bibfnamefont {J.~B.}\ \bibnamefont
  {Torrance}},\ }\bibfield  {title} {\bibinfo {title} {Model of the
  neutral-ionic phase transformation},\ }\href@noop {} {\bibfield  {journal}
  {\bibinfo  {journal} {Phys. Rev. Lett.}\ }\textbf {\bibinfo {volume} {47}},\
  \bibinfo {pages} {1750} (\bibinfo {year} {1981})}\BibitemShut {NoStop}%
\bibitem [{\citenamefont {Fabrizio}\ \emph {et~al.}(1999)\citenamefont
  {Fabrizio}, \citenamefont {Gogolin},\ and\ \citenamefont
  {Nersesyan}}]{ionichubbard1}%
  \BibitemOpen
  \bibfield  {author} {\bibinfo {author} {\bibfnamefont {M.}~\bibnamefont
  {Fabrizio}}, \bibinfo {author} {\bibfnamefont {A.~O.}\ \bibnamefont
  {Gogolin}},\ and\ \bibinfo {author} {\bibfnamefont {A.~A.}\ \bibnamefont
  {Nersesyan}},\ }\bibfield  {title} {\bibinfo {title} {From band insulator to
  mott insulator in one dimension},\ }\href@noop {} {\bibfield  {journal}
  {\bibinfo  {journal} {Phys. Rev. Lett.}\ }\textbf {\bibinfo {volume} {83}},\
  \bibinfo {pages} {2014} (\bibinfo {year} {1999})}\BibitemShut {NoStop}%
\bibitem [{\citenamefont {Garg}\ \emph {et~al.}(2006)\citenamefont {Garg},
  \citenamefont {Krishnamurthy},\ and\ \citenamefont
  {Randeria}}]{ionichubbard2}%
  \BibitemOpen
  \bibfield  {author} {\bibinfo {author} {\bibfnamefont {A.}~\bibnamefont
  {Garg}}, \bibinfo {author} {\bibfnamefont {H.~R.}\ \bibnamefont
  {Krishnamurthy}},\ and\ \bibinfo {author} {\bibfnamefont {M.}~\bibnamefont
  {Randeria}},\ }\bibfield  {title} {\bibinfo {title} {Can correlations drive a
  band insulator metallic?},\ }\href@noop {} {\bibfield  {journal} {\bibinfo
  {journal} {Phys. Rev. Lett.}\ }\textbf {\bibinfo {volume} {97}},\ \bibinfo
  {pages} {046403} (\bibinfo {year} {2006})}\BibitemShut {NoStop}%
\bibitem [{\citenamefont {Garg}\ \emph {et~al.}(2014)\citenamefont {Garg},
  \citenamefont {Krishnamurthy},\ and\ \citenamefont
  {Randeria}}]{ionichubbard3}%
  \BibitemOpen
  \bibfield  {author} {\bibinfo {author} {\bibfnamefont {A.}~\bibnamefont
  {Garg}}, \bibinfo {author} {\bibfnamefont {H.~R.}\ \bibnamefont
  {Krishnamurthy}},\ and\ \bibinfo {author} {\bibfnamefont {M.}~\bibnamefont
  {Randeria}},\ }\bibfield  {title} {\bibinfo {title} {Doping a correlated band
  insulator: A new route to half-metallic behavior},\ }\href@noop {} {\bibfield
   {journal} {\bibinfo  {journal} {Phys. Rev. Lett.}\ }\textbf {\bibinfo
  {volume} {112}},\ \bibinfo {pages} {106406} (\bibinfo {year}
  {2014})}\BibitemShut {NoStop}%
\bibitem [{\citenamefont {Bag}\ \emph {et~al.}(2015)\citenamefont {Bag},
  \citenamefont {Garg},\ and\ \citenamefont {Krishnamurthy}}]{ionichubbard4}%
  \BibitemOpen
  \bibfield  {author} {\bibinfo {author} {\bibfnamefont {S.}~\bibnamefont
  {Bag}}, \bibinfo {author} {\bibfnamefont {A.}~\bibnamefont {Garg}},\ and\
  \bibinfo {author} {\bibfnamefont {H.~R.}\ \bibnamefont {Krishnamurthy}},\
  }\bibfield  {title} {\bibinfo {title} {Phase diagram of the half-filled ionic
  hubbard model},\ }\href@noop {} {\bibfield  {journal} {\bibinfo  {journal}
  {Phys. Rev. B}\ }\textbf {\bibinfo {volume} {91}},\ \bibinfo {pages} {235108}
  (\bibinfo {year} {2015})}\BibitemShut {NoStop}%
\bibitem [{\citenamefont {Bag}\ \emph {et~al.}(2021)\citenamefont {Bag},
  \citenamefont {Garg},\ and\ \citenamefont {Krishnamurthy}}]{ionichubbard5}%
  \BibitemOpen
  \bibfield  {author} {\bibinfo {author} {\bibfnamefont {S.}~\bibnamefont
  {Bag}}, \bibinfo {author} {\bibfnamefont {A.}~\bibnamefont {Garg}},\ and\
  \bibinfo {author} {\bibfnamefont {H.~R.}\ \bibnamefont {Krishnamurthy}},\
  }\bibfield  {title} {\bibinfo {title} {Correlation driven metallic and
  half-metallic phases in a band insulator},\ }\href@noop {} {\bibfield
  {journal} {\bibinfo  {journal} {Phys. Rev. B}\ }\textbf {\bibinfo {volume}
  {103}},\ \bibinfo {pages} {155132} (\bibinfo {year} {2021})}\BibitemShut
  {NoStop}%
\bibitem [{\citenamefont {Chattopadhyay}\ \emph {et~al.}(2021)\citenamefont
  {Chattopadhyay}, \citenamefont {Krishnamurthy},\ and\ \citenamefont
  {Garg}}]{ionichubbard6}%
  \BibitemOpen
  \bibfield  {author} {\bibinfo {author} {\bibfnamefont {A.}~\bibnamefont
  {Chattopadhyay}}, \bibinfo {author} {\bibfnamefont {H.~R.}\ \bibnamefont
  {Krishnamurthy}},\ and\ \bibinfo {author} {\bibfnamefont {A.}~\bibnamefont
  {Garg}},\ }\bibfield  {title} {\bibinfo {title} {{Unconventional
  superconductivity in a strongly correlated band-insulator without doping}},\
  }\href@noop {} {\bibfield  {journal} {\bibinfo  {journal} {SciPost Phys.
  Core}\ }\textbf {\bibinfo {volume} {4}},\ \bibinfo {pages} {9} (\bibinfo
  {year} {2021})}\BibitemShut {NoStop}%
\bibitem [{\citenamefont {Hubbard}\ and\ \citenamefont
  {Flowers}(1963)}]{hubbard1}%
  \BibitemOpen
  \bibfield  {author} {\bibinfo {author} {\bibfnamefont {J.}~\bibnamefont
  {Hubbard}}\ and\ \bibinfo {author} {\bibfnamefont {B.~H.}\ \bibnamefont
  {Flowers}},\ }\bibfield  {title} {\bibinfo {title} {Electron correlations in
  narrow energy bands},\ }\href@noop {} {\bibfield  {journal} {\bibinfo
  {journal} {Proceedings of the Royal Society of London. Series A. Mathematical
  and Physical Sciences}\ }\textbf {\bibinfo {volume} {276}},\ \bibinfo {pages}
  {238} (\bibinfo {year} {1963})}\BibitemShut {NoStop}%
\bibitem [{\citenamefont {LeBlanc}\ \emph {et~al.}(2015)\citenamefont
  {LeBlanc}, \citenamefont {Antipov}, \citenamefont {Becca}, \citenamefont
  {Bulik}, \citenamefont {Chan}, \citenamefont {Chung}, \citenamefont {Deng},
  \citenamefont {Ferrero}, \citenamefont {Henderson}, \citenamefont
  {Jim\'enez-Hoyos}, \citenamefont {Kozik}, \citenamefont {Liu}, \citenamefont
  {Millis}, \citenamefont {Prokof'ev}, \citenamefont {Qin}, \citenamefont
  {Scuseria}, \citenamefont {Shi}, \citenamefont {Svistunov}, \citenamefont
  {Tocchio}, \citenamefont {Tupitsyn}, \citenamefont {White}, \citenamefont
  {Zhang}, \citenamefont {Zheng}, \citenamefont {Zhu},\ and\ \citenamefont
  {Gull}}]{hubbard2}%
  \BibitemOpen
  \bibfield  {author} {\bibinfo {author} {\bibfnamefont {J.~P.~F.}\
  \bibnamefont {LeBlanc}}, \bibinfo {author} {\bibfnamefont {A.~E.}\
  \bibnamefont {Antipov}}, \bibinfo {author} {\bibfnamefont {F.}~\bibnamefont
  {Becca}}, \bibinfo {author} {\bibfnamefont {I.~W.}\ \bibnamefont {Bulik}},
  \bibinfo {author} {\bibfnamefont {G.~K.-L.}\ \bibnamefont {Chan}}, \bibinfo
  {author} {\bibfnamefont {C.-M.}\ \bibnamefont {Chung}}, \bibinfo {author}
  {\bibfnamefont {Y.}~\bibnamefont {Deng}}, \bibinfo {author} {\bibfnamefont
  {M.}~\bibnamefont {Ferrero}}, \bibinfo {author} {\bibfnamefont {T.~M.}\
  \bibnamefont {Henderson}}, \bibinfo {author} {\bibfnamefont {C.~A.}\
  \bibnamefont {Jim\'enez-Hoyos}}, \bibinfo {author} {\bibfnamefont
  {E.}~\bibnamefont {Kozik}}, \bibinfo {author} {\bibfnamefont {X.-W.}\
  \bibnamefont {Liu}}, \bibinfo {author} {\bibfnamefont {A.~J.}\ \bibnamefont
  {Millis}}, \bibinfo {author} {\bibfnamefont {N.~V.}\ \bibnamefont
  {Prokof'ev}}, \bibinfo {author} {\bibfnamefont {M.}~\bibnamefont {Qin}},
  \bibinfo {author} {\bibfnamefont {G.~E.}\ \bibnamefont {Scuseria}}, \bibinfo
  {author} {\bibfnamefont {H.}~\bibnamefont {Shi}}, \bibinfo {author}
  {\bibfnamefont {B.~V.}\ \bibnamefont {Svistunov}}, \bibinfo {author}
  {\bibfnamefont {L.~F.}\ \bibnamefont {Tocchio}}, \bibinfo {author}
  {\bibfnamefont {I.~S.}\ \bibnamefont {Tupitsyn}}, \bibinfo {author}
  {\bibfnamefont {S.~R.}\ \bibnamefont {White}}, \bibinfo {author}
  {\bibfnamefont {S.}~\bibnamefont {Zhang}}, \bibinfo {author} {\bibfnamefont
  {B.-X.}\ \bibnamefont {Zheng}}, \bibinfo {author} {\bibfnamefont
  {Z.}~\bibnamefont {Zhu}},\ and\ \bibinfo {author} {\bibfnamefont
  {E.}~\bibnamefont {Gull}} (\bibinfo {collaboration} {Simons Collaboration on
  the Many-Electron Problem}),\ }\bibfield  {title} {\bibinfo {title}
  {Solutions of the two-dimensional hubbard model: Benchmarks and results from
  a wide range of numerical algorithms},\ }\href@noop {} {\bibfield  {journal}
  {\bibinfo  {journal} {Phys. Rev. X}\ }\textbf {\bibinfo {volume} {5}},\
  \bibinfo {pages} {041041} (\bibinfo {year} {2015})}\BibitemShut {NoStop}%
\bibitem [{\citenamefont {Plimpton}(1995)}]{lammps}%
  \BibitemOpen
  \bibfield  {author} {\bibinfo {author} {\bibfnamefont {S.}~\bibnamefont
  {Plimpton}},\ }\bibfield  {title} {\bibinfo {title} {Fast parallel algorithms
  for short-range molecular dynamics},\ }\href@noop {} {\bibfield  {journal}
  {\bibinfo  {journal} {Journal of Computational Physics}\ }\textbf {\bibinfo
  {volume} {117}},\ \bibinfo {pages} {1 } (\bibinfo {year} {1995})}\BibitemShut
  {NoStop}%
\bibitem [{\citenamefont {Stillinger}\ and\ \citenamefont {Weber}(1985)}]{sw}%
  \BibitemOpen
  \bibfield  {author} {\bibinfo {author} {\bibfnamefont {F.~H.}\ \bibnamefont
  {Stillinger}}\ and\ \bibinfo {author} {\bibfnamefont {T.~A.}\ \bibnamefont
  {Weber}},\ }\bibfield  {title} {\bibinfo {title} {Computer simulation of
  local order in condensed phases of silicon},\ }\href@noop {} {\bibfield
  {journal} {\bibinfo  {journal} {Phys. Rev. B}\ }\textbf {\bibinfo {volume}
  {31}},\ \bibinfo {pages} {5262} (\bibinfo {year} {1985})}\BibitemShut
  {NoStop}%
\bibitem [{\citenamefont {Naik}\ \emph {et~al.}(2019)\citenamefont {Naik},
  \citenamefont {Maity}, \citenamefont {Maiti},\ and\ \citenamefont
  {Jain}}]{kc_naik}%
  \BibitemOpen
  \bibfield  {author} {\bibinfo {author} {\bibfnamefont {M.~H.}\ \bibnamefont
  {Naik}}, \bibinfo {author} {\bibfnamefont {I.}~\bibnamefont {Maity}},
  \bibinfo {author} {\bibfnamefont {P.~K.}\ \bibnamefont {Maiti}},\ and\
  \bibinfo {author} {\bibfnamefont {M.}~\bibnamefont {Jain}},\ }\bibfield
  {title} {\bibinfo {title} {Kolmogorov–crespi potential for multilayer
  transition-metal dichalcogenides: Capturing structural transformations in
  moiré superlattices},\ }\href@noop {} {\bibfield  {journal} {\bibinfo
  {journal} {The Journal of Physical Chemistry C}\ }\textbf {\bibinfo {volume}
  {123}},\ \bibinfo {pages} {9770} (\bibinfo {year} {2019})}\BibitemShut
  {NoStop}%
\bibitem [{\citenamefont {Soler}\ \emph {et~al.}(2002)\citenamefont {Soler},
  \citenamefont {Artacho}, \citenamefont {Gale}, \citenamefont {Garc{\'{\i}}a},
  \citenamefont {Junquera}, \citenamefont {Ordej{\'{o}}n},\ and\ \citenamefont
  {S{\'{a}}nchez-Portal}}]{siesta}%
  \BibitemOpen
  \bibfield  {author} {\bibinfo {author} {\bibfnamefont {J.~M.}\ \bibnamefont
  {Soler}}, \bibinfo {author} {\bibfnamefont {E.}~\bibnamefont {Artacho}},
  \bibinfo {author} {\bibfnamefont {J.~D.}\ \bibnamefont {Gale}}, \bibinfo
  {author} {\bibfnamefont {A.}~\bibnamefont {Garc{\'{\i}}a}}, \bibinfo {author}
  {\bibfnamefont {J.}~\bibnamefont {Junquera}}, \bibinfo {author}
  {\bibfnamefont {P.}~\bibnamefont {Ordej{\'{o}}n}},\ and\ \bibinfo {author}
  {\bibfnamefont {D.}~\bibnamefont {S{\'{a}}nchez-Portal}},\ }\bibfield
  {title} {\bibinfo {title} {The {SIESTA} method forab initioorder-nmaterials
  simulation},\ }\href@noop {} {\bibfield  {journal} {\bibinfo  {journal}
  {Journal of Physics: Condensed Matter}\ }\textbf {\bibinfo {volume} {14}},\
  \bibinfo {pages} {2745} (\bibinfo {year} {2002})}\BibitemShut {NoStop}%
\bibitem [{\citenamefont {Alden}\ \emph {et~al.}(2013)\citenamefont {Alden},
  \citenamefont {Tsen}, \citenamefont {Huang}, \citenamefont {Hovden},
  \citenamefont {Brown}, \citenamefont {Park}, \citenamefont {Muller},\ and\
  \citenamefont {McEuen}}]{op1}%
  \BibitemOpen
  \bibfield  {author} {\bibinfo {author} {\bibfnamefont {J.~S.}\ \bibnamefont
  {Alden}}, \bibinfo {author} {\bibfnamefont {A.~W.}\ \bibnamefont {Tsen}},
  \bibinfo {author} {\bibfnamefont {P.~Y.}\ \bibnamefont {Huang}}, \bibinfo
  {author} {\bibfnamefont {R.}~\bibnamefont {Hovden}}, \bibinfo {author}
  {\bibfnamefont {L.}~\bibnamefont {Brown}}, \bibinfo {author} {\bibfnamefont
  {J.}~\bibnamefont {Park}}, \bibinfo {author} {\bibfnamefont {D.~A.}\
  \bibnamefont {Muller}},\ and\ \bibinfo {author} {\bibfnamefont {P.~L.}\
  \bibnamefont {McEuen}},\ }\bibfield  {title} {\bibinfo {title} {Strain
  solitons and topological defects in bilayer graphene},\ }\href@noop {}
  {\bibfield  {journal} {\bibinfo  {journal} {Proceedings of the National
  Academy of Sciences}\ }\textbf {\bibinfo {volume} {110}},\ \bibinfo {pages}
  {11256} (\bibinfo {year} {2013})}\BibitemShut {NoStop}%
\bibitem [{\citenamefont {Gargiulo}\ and\ \citenamefont {Yazyev}(2017)}]{op2}%
  \BibitemOpen
  \bibfield  {author} {\bibinfo {author} {\bibfnamefont {F.}~\bibnamefont
  {Gargiulo}}\ and\ \bibinfo {author} {\bibfnamefont {O.~V.}\ \bibnamefont
  {Yazyev}},\ }\bibfield  {title} {\bibinfo {title} {Structural and electronic
  transformation in low-angle twisted bilayer graphene},\ }\href@noop {}
  {\bibfield  {journal} {\bibinfo  {journal} {2D Materials}\ }\textbf {\bibinfo
  {volume} {5}},\ \bibinfo {pages} {015019} (\bibinfo {year}
  {2017})}\BibitemShut {NoStop}%
\bibitem [{\citenamefont {Mermin}(1979)}]{aster1}%
  \BibitemOpen
  \bibfield  {author} {\bibinfo {author} {\bibfnamefont {N.~D.}\ \bibnamefont
  {Mermin}},\ }\bibfield  {title} {\bibinfo {title} {The topological theory of
  defects in ordered media},\ }\href@noop {} {\bibfield  {journal} {\bibinfo
  {journal} {Rev. Mod. Phys.}\ }\textbf {\bibinfo {volume} {51}},\ \bibinfo
  {pages} {591} (\bibinfo {year} {1979})}\BibitemShut {NoStop}%
\end{thebibliography}
\end{document}